# Dissociation of dark matter and gas in cosmic large-scale structure

William McDonald ⋆, Danail Obreschkow ⋆ and Lilian Garratt-Smithson
*International Centre for Radio Astronomy Research, M468, University of Western Australia, 35 Stirling Hwy, Perth, WA 6009, Australia*

**ABSTRACT**

The partial spatial separation of cold dark matter (DM) and gas is a ubiquitous feature in the formation of cosmic large-scale structure. This separation, termed dissociation, is prominent in galaxy clusters that formed through collisions of massive progenitors, such as the famous 'Bullet' cluster. A direct comparison of the incidence of such dissociated structures with theoretical predictions is challenged by the rarity of strongly dissociated systems and the difficulty to quantify dissociation. This paper introduces a well-defined dimension-less dissociation index $S \in [-1, 1]$ that encodes the quadrupole difference between DM and gas in a custom region. Using a simulation of cosmic large-scale structure with cold DM and ideal non-radiating gas, in $\Lambda$CDM cosmology, we find that 90 per cent of the haloes are positively dissociated ($S > 0$), meaning their DM is more elongated than their gas. The spatial density of highly dissociated massive structures appears consistent with observations. Through idealized $N$-body + SPH simulations of colliding gaseous DM haloes, we further explore the details of how ram-pressure causes dissociation in binary collisions. A suite of 300 such simulations reveals a scale-free relation between the orbital parameters of binary collisions and the resulting dissociation. Building on this relation, we conclude that the frequency of dissociated structures in non-radiative cosmological simulations is nearly fully accounted for by the major (mass ratio >1:10) binary collisions predicted by such simulations. In principle, our results allow us to constrain the orbital parameters that produced specific observed dissociated clusters.

## 1 INTRODUCTION

In the standard cold dark matter model with a cosmological constant $\Lambda$ ($\Lambda$CDM), a fluid of dark matter (DM) and gas gradually clumps into haloes that grow hierarchically, through the accretion of other haloes (Peebles 1965; Davis et al. 1985) and possibly unbound material. Galaxies form inside the gravitational potential wells of these haloes (White & Rees 1978). As DM haloes coalesce, their galaxies are brought together, leading to galaxy mergers, galaxy groups, and clusters. Galaxy clusters are the largest virialized objects in the observable Universe (for a relevant review of galaxy cluster formation, see Kravtsov & Borgani 2012). Such clusters typically consist of a massive (> $10^{14}$ M$_\odot$) host DM halo, a relatively massive central elliptical galaxy, surrounded by satellite galaxies and permeated by a significant intracluster medium (ICM).

Interactions between cluster-scale haloes are expected to be common occurrences in the recent history of clusters, as they tend to be dynamically young systems (e.g. Cohn & White 2005; Fakhouri, Ma & Boylan-Kolchin 2010). Major mergers between clusters are easily observable because they inject substantial amounts of energy into the ICM (e.g. Ricker & Sarazin 2001; Sarazin 2002). The morphology of a galaxy cluster and the structure of a DM halo depends on the assembly history (e.g. White & Frenk 1991; Ludlow et al. 2013; Correa et al. 2015; Drakos et al. 2019a, b), specifically cluster–cluster mergers can lead to the formation of many observable transient structures within a galaxy cluster (e.g. Roettiger, Burns & Loken 1996; Poole et al. 2006; Markevitch & Vikhlinin 2007; Golovich et al. 2019b).

A well-known and scientifically important case of an ongoing collision between two galaxy clusters is 1ES−0657558 (first studied by Tucker, Tananbaum & Remillard 1995; Tucker et al. 1998), the 'Bullet' cluster. Combined X-ray observations and optical gravitational lensing studies show that the Bullet cluster is dominated by two massive DM sub-haloes, which flank a centralized concentration of an incredibly hot, X-ray emitting ICM. A litany of idealized simulations agree that this peculiar morphology and the dynamic state of the Bullet cluster can be formed in a $\Lambda$CDM setting (e.g. Takizawa 2005, 2006; Milosavljević et al. 2007; Springel & Farrar 2007; Mastropietro & Burkert 2008; Dawson 2013; Lage & Farrar 2014), confirming the idea that the Bullet is the smoking gun of a near head-on collision of two massive progenitor clusters (e.g. Markevitch et al. 2002; Clowe, Gonzalez & Markevitch 2004; Bradač et al. 2006; Clowe et al. 2006). As the two progenitors collided, the non-collisional components (DM and galaxies) moved past one another. In turn, the gas was subject to dynamic pressure, creating the obvious spatial offset between the galaxies and DM components from the ICM. Systems such as the Bullet cluster are therefore useful laboratories for studying the (self-)interaction of DM and gas (e.g. Markevitch et al. 2004; Randall et al. 2008).

The Bullet cluster is not unique. There are numerous observations of galaxy clusters which exhibit a significant spatial separation between the ICM and DM components. Some examples of such clusters are Abell 520 (Mahdavi et al. 2007; Jee et al. 2012), MACS J0025.4-1222 (Baby Bullet Cluster; Bradač et al. 2008), DLSCL

---

⋆ E-mail: william.mcdonald@icrar.org (WM); danail.obreschkow@icrar.org (DO)

J0916.2+2951 (Musket Ball Cluster; Dawson et al. 2012), SL2S J08544-0121 (Bullet Group; Gastaldello et al. 2014), ACT-CL J0102-4915 (El Gordo Jee et al. 2014; Ng et al. 2015; Diego et al. 2020), MACS J1149.5+2223 (Golovich et al. 2016), ZwCl 0008.8+5215 (Golovich et al. 2017), and Abell 2034 (Monteiro-Oliveira et al. 2018). Much like the Bullet, many of these clusters have been used as probes for the nature of dark matter, specifically the collisional cross-section, (e.g. Bradač et al. 2008; Clowe et al. 2012; Dawson et al. 2012; Gastaldello et al. 2014; Kahlhoefer et al. 2014; Harvey et al. 2015; Kim, Peter & Wittman 2017, for an overview see Tulin & Yu 2018) and various behaviours of the ICM under relatively extreme conditions (e.g. Van Weeren et al. 2017; Ha, Ryu & Kang 2018; Lourenço et al. 2020).

The spatial separation of gas and DM observed in clusters is commonly known as 'dissociation' (Dawson et al. 2012), and the mergers/collisions causing the separation are called 'dissociative' mergers/collisions. Explicitly, Dawson et al. (2012) defines a dissociative merger as (1) a merger between two similarly massive clusters, (2) with a small impact parameter, (3) that is observed while the ICM is significantly offset from the galaxies and DM, and (4) occurs roughly perpendicular to the observer's line of sight, such that the dissociation is apparent. In most of this paper, our definition of 'dissociation' is less restrictive and independent of the observers' position. Of course, the ability to detect dissociated structures in observations depends on a system's inclination relative to the line of sight.

The presence of a dissociative morphology does not necessitate a merger, for example in hyperbolic collisions (positive orbital energy), most of the system will not eventually merge. A spatial separation between the ICM and DM can be observed regardless of whether or not a merger follows a collision.

The level of dissociation caused by a merger/collision depends on the characteristics of the progenitor haloes and their orbital parameters. However, quantifying this level of dissociation requires a consistent formal definition of dissociation in two-fluid systems (here DM and gas). To date most of the discussion surrounding the dissociation of galaxy clusters has largely been focused on replicating specific geometries, accurately mapping the dynamics of a cluster (see Golovich et al. (2019a, b) for an overview of detecting and analysing merging clusters) and identifying analogues of particular observations (as seen in the extensive discussion on the existence of bullet-like clusters in a ΛCDM paradigm; Hayashi & White 2006; Forero-Romero, Gottlöber & Yepes 2010; Lee & Komatsu 2010; Fernández-Trincado et al. 2014; Watson et al. 2014; Bouillot et al. 2015; Kraljic & Sarkar 2015; Lage & Farrar 2015; Thompson, Davé & Nagamine 2015). The continued expansion of X-ray surveys (such as the remaining planned full-sky surveys from *eROSITA*; Predehl et al. 2021) promises to dramatically increase our ability to observe dissociated structures, thus a quantitative measure and reproducible formal definition is timely.

In this paper, we introduce the dissociation index, $S \in [-1, 1]$, a dimension-less parameter that quantifies the level of quadrupole separation between two fluids, such as the DM and gas in a halo. Using this index, we then study the distribution of dissociation in cosmological simulations of large-scale structure formation, as well as in idealized simulations of binary halo–halo encounters. Combining these complementary simulations, we conclude that the large-scale dissociation between gas and DM in the absence of galaxy physics is almost entirely accounted for by major (mass ratio >1:10) binary halo collisions.

We devote Section 2 of this paper to the definition and interpretation of the dissociation index. A direct application is presented in Section 3, where we study the level and frequency of disso-ciation in haloes in a cosmological simulation from the SURFS simulation suite. In Section 4, we describe a series of idealized *N*-body+Hydrodynamic simulations of binary collisions between gaseous DM haloes and discuss the resulting dissociation as a function of the binary orbits' initial conditions. Section 5 furthers this discussion using the statistics of halo–halo collisions of cosmic large-scale structure. Finally Section 6 summarizes the key findings of this paper and potential ways in which the dissociation index can be utilized in studying large-scale structure formation.

## 2 QUANTIFYING THE DISSOCIATION OF GASEOUS HALOES

In this section, we introduce the dissociation index *S*, a dimensionless parameter to quantify the spatial separation between two fluids in terms of their quadrupole difference.

### 2.1 Definition of the dissociation index

The general phenomenology of strongly dissociated clusters (see examples in Section 1) is such that the DM is elongated into a prolate, 'dumbbell-like' shape, whereas the ICM tends to be more concentrated at the centre of the structure. This qualitative geometry is expected in the aforementioned formation scenario of two colliding gaseous haloes with a non-collisional DM component. Dissociated systems are thus characterized by a stronger quadrupole in DM than gas. A useful way of quantifying these quadrupoles is to expand the mean surface of the DM and gas, individually, into spherical harmonics.

Formally, the quadrupole of the mean surface of a mass distribution $\rho(\boldsymbol{x})$, about the origin $\boldsymbol{x} = 0$, is fully characterized via the five quadrupole coefficients,

$$f_m = \frac{1}{M} \int d^3\boldsymbol{x}\, r\, \rho(\boldsymbol{x})\, Y_2^m(\boldsymbol{x}), \tag{1}$$

where $M = \int d^3\boldsymbol{x}\, \rho(\boldsymbol{x})$ is the total mass, $r = |\boldsymbol{x}|$ is the radial coordinate, and $Y_2^m$ are the spherical harmonic functions of degree $l = 2$ (quadrupole) and order $m = -2, -1, 0, 1, 2$. The total quadrupole amplitude is then given by the norm

$$q = \left(\sum_{m=-2}^{2} |f_m|^2\right)^{1/2}. \tag{2}$$

This norm is independent of the choice of the spherical harmonics basis, as long as it is orthonormal, $\iint d\Omega\, Y_l^m Y_{l'}^{m'*} = \delta_{ll'}\delta_{mm'}$. In particular, it does not matter whether real or complex spherical harmonics are used.[1] Either way, the orthonormalization conditions imply that $q$ is bound between 0 and $\sqrt{5/(4\pi)}\,\bar{r}$, where

$$\bar{r} = \frac{1}{M} \int d^3\boldsymbol{x}\, r\, \rho(\boldsymbol{x}) \tag{3}$$

is the mean radius.

We can now quantify the quadrupole-like dissociation between two fluids – let us call them '+' and '−' to remain generic – via the difference in their respective quadrupole amplitudes. The bounds of these amplitudes make it natural to normalize this difference as

$$S = \sqrt{\frac{4\pi}{5}} \frac{q_+ - q_-}{\bar{r}_{\rm max}}, \quad (4)$$

where $\bar{r}_{\rm max} = \max\{\bar{r}_+, \bar{r}_-\}$ is the larger of the two mean radii. Given this normalization, $S$ is bound to the interval $[-1, 1]$.

---

[1]For reference, if $\boldsymbol{x} = (x, y, z)$ and $r = |\boldsymbol{x}|$, the complex spherical harmonics, $Y_2^m(\boldsymbol{x})$, in the Condon-Shortley phase convention, are $Y_2^{\pm 2} = \sqrt{15/(32\pi)} \cdot (x \pm iy)^2/r^2$, $Y_2^{\pm 1} = \mp\sqrt{15/(8\pi)} \cdot (x \pm iy)z/r^2$, $Y_2^0 = \sqrt{5/(16\pi)} \cdot (3z^2 - r^2)/r^2$. The corresponding real basis functions, $Y_{lm}(\boldsymbol{x})$, are $Y_{2,-2} = \sqrt{15/(4\pi)} \cdot xy/r^2$, $Y_{2,-1} = \sqrt{15/(4\pi)} \cdot yz/r^2$, $Y_{2,0} = Y_2^0$, $Y_{2,1} = \sqrt{15/(4\pi)} \cdot xz/r^2$, $Y_{2,2}\sqrt{15/(16\pi)} \cdot (x^2 - y^2)/r^2$.

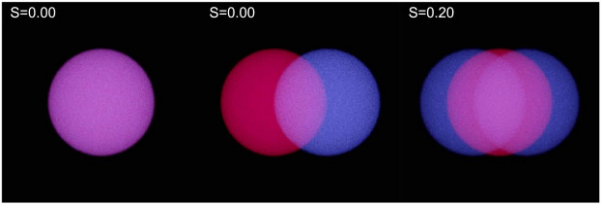

**Figure 1.** In each panel of this figure, the dissociation index is shown (top left-hand side) for a series of identical spheres which illustrate various idealized galaxy clusters composed of DM (blue) and an ICM (red). The overlapping regions are coloured purple. Left-hand panel: two identical spheres of an ICM and DM share the same geometric centre, forming a monopole, thus $q_{\rm DM} = q_{\rm gas} = 0$. Centre: the same two identical spheres are offset from one another, resulting in a dipole-like morphology. Right-hand panel: an idealized dissociated galaxy cluster is depicted by mirroring the DM spatial distribution of the centre panel about the vertical axis. The DM resembles a quadrupole ($q_{\rm DM} > 0$) whilst the ICM remains a monopole ($q_{\rm gas} = 0$), thus $S > 0$.

A greater value of $|S|$ means that one fluid has a more asymmetrical spatial distribution relative to the other and the sign of $S$ indicates which fluid has the dominant quadrupole. The extremes of $S = \pm 1$ are reached in the situation where the fluid + or −, respectively, is completely dissociated into two antipodal points, whereas the other fluid is spherically distributed at the centre. Henceforth, the parameter $S$ is referred to as the 'dissociation index'.

The remaining piece in the definition of $S$ is a definition of the origin, $x = 0$. One might be tempted to set this origin equal to the combined centre of mass of the two fluids. However, this choice has the disadvantage that the value of $S$ would change if we only vary the mass ratio between the two fluids without changing their shapes and relative position. A straightforward way to avoid this dependence on the mass ratio is to define the origin as the average position between the centres of mass, $x_+^{\rm CM}$ and $x_-^{\rm CM}$, of the two fluids, respectively.

In this paper, we identify the fluid '+' with the DM and the fluid '−' with the gas, i.e. the ICM in the case of clusters. With this convention, typical dissociated structures, such as the Bullet cluster, have a positive dissociation index.

### 2.2 Illustration of the dissociation index

Figs 1 and 2 provide an intuition for the dissociation index $S$ using a set of toy geometries. Each figure consists of three panels, each containing a different arrangement of uniform spheres of DM (blue) and gas (red). Regions, where the two fluids overlap, appear purple.

The message of Fig. 1 is that $S$ differs from zero if and only if the DM and gas exhibit a quadrupole difference. In particular, a dipole-like displacement as shown (centre panel) does not affect the value of $S$. In turn, Fig. 2 illustrates how $S$ varies with the degree of the quadrupole difference from differing morphologies. The rightmost panels in both figures show the same situation with reversed substances (colours). This inversion corresponds to a sign-flip of $S$, as obvious from the antisymmetry between $q_+$ and $q_-$ in equation (2).

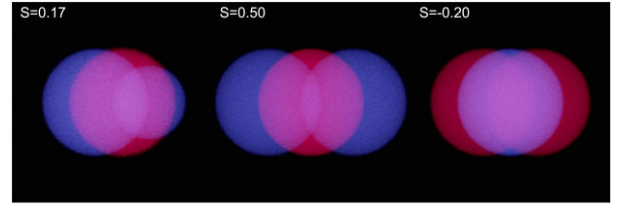

**Figure 2.** Each panel of this figure shows a permutation of an idealized dissociated cluster as established in the rightmost panel of Fig. 1. Left-hand panel: an idealized dissociated structure with DM lobes of unequal mass is shown by scaling the radii of the right DM sphere, i.e. mimicking a collision between progenitors with a 1:3 mass ratio. The dissociation index decreases ($S = 0.2 \to S = 0.17$) as the DM distribution is 'less' quadrupole like. Centre: shows the initial idealized dissociated cluster, with an increased level of dissociation ($S = 0.2 \to S = 0.5$). The two DM spheres are moved further from the geometric centre, increasing $q_{\rm DM}$. Right-hand panel: a negatively dissociated cluster is shown. All DM and ICM points of the initial idealized cluster have been swapped, inverting the sign of $S$ ($S = 0.2 \to S = -0.2$).

### 2.3 Calculating the dissociation index from particle data

In *N*-body simulations both the gas and DM can be represented by discrete particles. Neglecting the spatial smoothing of these particles used by many integrators, the mass densities $\rho(\boldsymbol{x})$ of the two fluids are a sum of Dirac delta functions. The general formalism of Section 2.1 then simplifies as follows.

First, we must translate the particle positions $\boldsymbol{X}_i$ of all considered DM and gas particles to their mean centre of mass,

$$\boldsymbol{X}_0 = \frac{1}{2M_{\rm DM}} \sum_{i=1}^{N_{\rm DM}} m_i^{\rm DM} \boldsymbol{X}_i^{\rm DM} + \frac{1}{2M_{\rm gas}} \sum_{i=1}^{N_{\rm gas}} m_i^{\rm gas} \boldsymbol{X}_i^{\rm gas}, \quad (5)$$

where $N_{\rm DM}$ and $N_{\rm gas}$ are the numbers of particles of each species, and $M_{\rm DM}$ and $M_{\rm gas}$ are their respective total masses. In the following, $\boldsymbol{x}_i = \boldsymbol{X}_i - \boldsymbol{X}_0$ denote the centred positions.

For both species (DM and gas), equation (1) simplifies to

$$f_m = \frac{1}{M} \sum_{i=1}^{N} r_i \, m_i \, Y_2^m(\boldsymbol{x}_i), \quad (6)$$

where $r_i = |\boldsymbol{x}_i|$ are the distances of the particles from the origin. Thus, the mean radius is

$$\bar{r} = \frac{1}{M} \sum_{i=1}^{N} r_i \, m_i. \quad (7)$$

Equations (6) and (7) need to be evaluated separately for gas and DM particles. The dissociation index is then given by equation (4) as,

$$S = \sqrt{\frac{4\pi}{5}} \frac{q_{\rm DM} - q_{\rm gas}}{\max\{\bar{r}_{\rm DM}, \bar{r}_{\rm gas}\}}, \quad (8)$$

where $q_{\rm DM}$ and $q_{\rm gas}$ are the quadrupole amplitudes computed from the coefficients $f_m$ of the two species, respectively, via equation (2).

### 2.4 Observational estimator of the dissociation index

In the quest to compare dissociation levels predicted by simulations with observations of dissociated clusters, we have to address the question of how the dissociation index, $S$, can be computed from observational data. The main difference to *N*-body data is that the position coordinate along the line-of-sight direction is missing in observations. Without loss of generality, we can identify this direction with the $z$-axis of a Cartesian coordinate system, such that the 2D image of a cluster is described in the $(x, y)$-plane. One might identify $x$ and $y$ with right ascension and declination, but this is not a requirement. The only necessary condition for the following is that $x$ and $y$ are orthogonal.

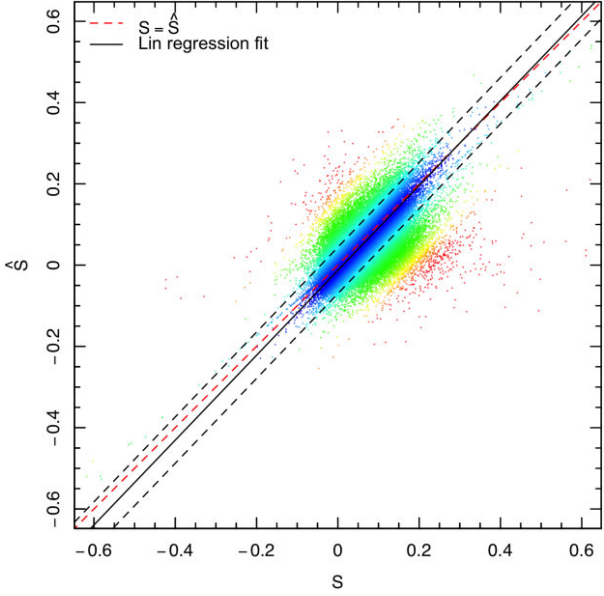

**Figure 3.** A comparison of $S$ and the observational proxy $\hat{S}$ computed for all well resolved haloes (see Section 3.2 for definition of well resolved haloes with respect to this paper) in the $N$-body + hydrodynamic cosmological simulation SURFS L210N1024NR (described in Section 3). The best linear regression fit (black line) out to one residual standard deviation (dashed black line) is overlaid and each point is coloured according to the number of standard deviations from the fitted line. A 1:1 relationship (dashed red line) is overlaid for comparison.

The consequence of missing the $z$-coordinate is apparent when considering the spherical harmonics as expressed in the footnote of Section 2.1. If the $z$-coordinate vanishes, both $Y_2^{-1}$ and $Y_2^1$ vanish, and $Y_2^0$ becomes a constant. Therefore, the three quadrupole coefficients $f_{-1}, f_0$ and $f_1$ cannot carry any information. Instead all the quadrupole information of the 2D data is encoded in $f_{-2}$ and $f_2$. Thus, our aim is to find an estimator, $\hat{S}$ of the true dissociation index $S$, using only these two quadrupole coefficients.

*2.4.1 Case of face-on collisions*

We first consider the case of a halo–halo collision observed face-on, i.e. such that the collision path lies in the plane of the sky, perpendicular to the line of sight. In this case, if the structure is assumed to be rotationally symmetric around the collision axis, it can be shown that simply ignoring $f_{-1}, f_0, f_1$ in computing the quadrupole amplitudes provides a remarkably good approximation of $S$ (as we will demonstrate later in Fig. 3).

To be explicit, let us assume that the empirical data of a dissociated cluster takes the form of two pixelated maps representing the gas column density (e.g. inferred from X-ray imaging) and DM column density (e.g. inferred from gravitational lensing). The estimated gas mass in each pixel $i$ is $m_i^{\rm gas}$; and likewise the DM mass is $m_i^{\rm DM}$.

As in the case of particle data (Section 2.3), we first translate our coordinates to the mean centre of mass, via equation (5), but summing over the pixels rather than particles and using this equation in two dimensions rather than three. Next, we set the three quadrupole coefficients with no information to zero, $f_{-1} = f_0 = f_1 = 0$. This simplifies equations (2) and (6) to

$$\tilde{q} = \sqrt{\frac{15}{16\pi}} \left( 4 \left[ \sum w_i x_i y_i \right]^2 + \left[ \sum w_i (x_i^2 - y_i^2) \right]^2 \right)^{1/2}, \quad (9)$$

where the sum goes over all pixels. The pixel weights $w_i$, are defined as $w_i = (m_i/M)/r_i$, where $r_i$ are now the 2D radii $r_i = (x_i^2 + y_i^2)^{1/2}$. The mean radius of each species is computed via equation (7), where the sum goes again over the pixels rather than particles.

As expected, equation (9) is invariant under uniform rotations, i.e. under any transformation $(x_i, y_i) \mapsto (x_i \cos\theta - y_i \sin\theta, y_i \cos\theta + x_i \sin\theta)$, as well as under mirroring operations.

Equations (9) and (7) need to be evaluated separately for each species (DM and gas). The estimator of the dissociation index is then

$$\hat{S} = \sqrt{\frac{4\pi}{5}} \frac{\tilde{q}_{\rm DM} - \tilde{q}_{\rm gas}}{\max\{\bar{r}_{\rm DM}, \bar{r}_{\rm gas}\}}. \quad (10)$$

Fig. 3 shows a comparison between the true value of $S$ computed from 3D particle data, and the estimator $\hat{S}$ computed only from $(x, y)$-maps after rotating the structures such that the quadrupole is maximized in these two dimensions. The points represent the friends-of-friends haloes in the non-radiative cosmological simulation that will be discussed in detail in Section 3. The black lines show a linear regression with 1-sigma standard deviations.

It is apparent from Fig. 3 that $\hat{S}$ (equation 10) is a nearly unbiased estimator of $S$. The root-mean-square deviation between the two is 0.06. In summary, if $\hat{S}$ is used as an observational proxy of $S$, one would expect a symmetrical statistical uncertainty with a standard deviation of about 0.06.

*2.4.2 Effects of inclination*

So far, we have assumed that the maximum quadrupole amplitude of a dissociated system lies in the plane of the sky. Real systems are likely to be differently oriented. To quantify this effect, we define the inclination $i \in [0°, 90°]$ as the (smaller) angle between the major axis of the DM quadrupole and the line of sight. The face-on situation discussed in Section 2.4.1 is characterized by $i = 90°$.

In general, the value of $\hat{S}$ decreases as the inclination decreases. Analytically, it can be shown, that, in the limit of small true dissociation indices ($|S| \ll 1$), the relation between the observed value $\hat{S}$ (equation 10) and the corresponding estimator $\hat{S}_{\rm corrected}$ of a face-on view is defined as

$$\hat{S} = \hat{S}_{\rm corrected} \sin^2 i. \quad (11)$$

Of course, for face-on systems, $\hat{S} = \hat{S}_{\rm corrected}$.

Numerically, we find that equation (11) is a reasonable approximation for systems with $|S| \lesssim 0.15$ which are roughly symmetric between the two DM lumps. A few such examples are shown in Fig. 4. The figure also includes some counterexamples, where equation (11) is quite inaccurate. The most extreme case is that with $S = 0.9$, where the $\hat{S}/\hat{S}_{\rm corrected}$–$i$ relation (blue line) differs considerably from equation (11). Note, however, that cluster-sized systems with dissociation indices above 0.3 are extremely rare in a $\Lambda$CDM universe (see Section 3). Additionally for every possible geometry relating to a given value of $S$, $\hat{S}$ is affected by inclination differently as seen in Fig. 4 by examples *d* and *f*. However, as a whole equation (11) is a reasonable proxy for inclination corrections, especially in view of the empirical challenge and systematic uncertainties associated with inclination measurements.

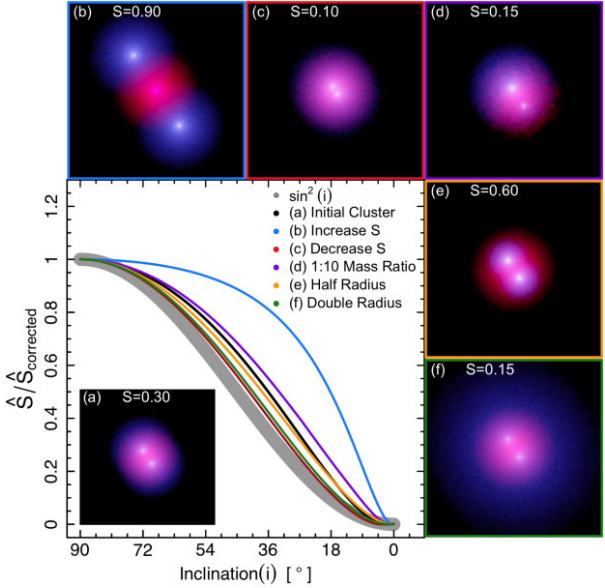

**Figure 4.** An illustration of $\hat{S}$ measured as a function of the inclination ($i$) for 2D projections of various idealized dissociated structures compared to the dissociation predicted by $\hat{S}_{corrected} \sin^2(i)$ (grey). (a): shows an initial dissociated cluster constructed of 'n' many point masses representing DM (blue) and the ICM (red) that is rationally symmetrical about the collision axis. The overlapping regions of this structure are shown as purple and column density is given by the colour intensity. The cluster is rotated about the axis orthogonal to both the collision axis and the line of sight, where $\hat{S}$ is computed at each angle of inclination, $i$. This is repeated for a variety of scenarios enacted on the DM of (a) as shown in; (b), (c), (d), (e), and (f), each of which maintain rotational symmetry about the collision axis. The true dissociation indices, $S$, of each idealized scenario are given in the top left of each panel.

*2.4.3 Example of the bullet cluster*

Let us finally apply the formalism of this section to estimate the dissociation index of the famous Bullet cluster: $S_{Bullet}$ (see Section 1). We assume that this colliding system is seen face-on ($i = 90°$) and has rotational symmetry about the collision axis so that we can use the formalism of Section 2.4.1 to estimate $S_{Bullet}$ without applying the inclination correction of equation (11).

We adopt the gas and DM density maps presented in Clowe et al. (2006), which were derived from X-ray temperature and gravitational lensing maps. Doing so, our estimator is $\hat{S}_{Bullet} = 0.335$, hence the true index is determined to be $S_{Bullet} = 0.335 \pm 0.06$.

## 3 DISSOCIATION IN A COSMOLOGICAL CONTEXT

Using the dissociation index $S$ (Section 2) it is now possible to quantify the expected dissociation of haloes in a $\Lambda$CDM universe. To do so, we evaluate $S$ for all the friends-of-friends haloes found in a cosmological simulation with cold DM and an ideal (non-radiating) gas from the Synthetic UniveRes For Surveys (SURFS) simulations suite (Elahi et al. 2018).

### 3.1 The cosmological simulation

SURFS is a collection of $N$-body and $N$-body + smoothed particle hydrodynamics (SPH) simulations run on a memory-lean version of GADGET-2 (Springel 2005) that assumes a $\Lambda$CDM cosmology. All these runs rely on Planck 2016 cosmological parameters (Planck Collaboration XIII 2016) with normalized redshift $z = 0$ densities of $\Omega_\Lambda = 0.6879$ (dark energy), $\Omega_M = 0.3121$ (all matter), and $\Omega_b = 0.0491$ (baryonic matter). The Hubble constant is $H_0 = 100h \text{km s}^{-1} \text{Mpc}^{-1}$ with dimension-less Hubble parameter $h = 0.6751$. The scalar spectral index is $n_s = 0.96553$ and the power spectrum normalization $\sigma_8 = 0.8150$.

Each simulation in SURFS evolves particles from initial conditions at redshift $z = 24$ to redshift $z = 0$, where the initial conditions were generated via the second-order Lagrangian perturbation theory (Crocce, Pueblas & Scoccimarro 2012) and a matter transfer function generated by CAMB (Lewis, Challinor & Lasenby 2000). Each simulation produces 200 snapshots, at evenly spaced intervals in the logarithm of the scale factor. Halo catalogues are created at each snapshot via VELOCIraptor (Elahi et al. 2019a), a structure finder using a 3D friends-of-friends (3DFOF) algorithm (Davis et al. 1985) to identify haloes followed by a 6DFOF algorithm to further identify substructure.

A particular SURFS run considered in this work includes an ideal gas in addition to cold DM. This $N$-body + SPH simulation (internally referred to as L210N1024NR, see table 1 in Elahi et al. 2018) uses a co-moving cubic box with a side length of $210\,h^{-1}$ cMpc, with periodic boundary conditions. This box contains $2 \times 1024^3$ particles (half DM, half gas) with constant masses of $6.29 \times 10^8\,h^{-1}\,M_\odot$ (DM) and $1.17 \times 10^8\,h^{-1}\,M_\odot$ (gas), respectively, and a gravitational a softening length of $\epsilon = 6.8\,h^{-1}$ ckpc.

In the real Universe, only massive, cluster-scale haloes ($\gtrsim 10^{14}\,M_\odot$) can maintain a significant hot gas component, whereas smaller, galaxy-scale haloes cool efficiently. However, DM simulations with an ideal gas (without radiative cooling) preserve the scale-free behaviour of pure DM runs. Therefore, the gaseous haloes in L210N1024NR are (nearly) self-similar across all masses. In particular, even haloes whose mass is too low ($\lesssim 10^{12}\,M_\odot$) to maintain a warm/hot atmosphere in the real Universe, maintain an equally significant halo gas component as cluster-scale haloes in non-radiative simulations. We can exploit this scale invariance and use the many sub-cluster-scale haloes as proxies of cluster-scale haloes.

### 3.2 Dissociation indices of a cosmological simulation

To determine the statistical distribution of $S$ in L210N1024NR, we select all the first-generation haloes more massive than $7.5 \times 10^{11}\,M_\odot$, which roughly corresponds to $\gtrsim 1000$ particles per species, depending on the exact ratio between gas and DM particles. This cut was applied because haloes with fewer particles are subject to significant Poisson noise that add random numerical aberrations to $S$ (see illustration in Appendix A). Moreover, small numbers of particles make it more difficult for FOF algorithms to robustly identify structures (e.g. Knebe et al. 2011, 2013), additionally summarized by Thompson et al. 2015; Angulo & Hahn 2022). In this paper, we refer to haloes in L210N1024NR that are $< 7.5 \times 10^{11}\,M_\odot$ as being poorly resolved and the remaining more massive haloes as being well resolved when computing $S$.

At redshift $z = 0$, our sample has a total of $\approx 70\,500$ haloes above the mass cut of $7.5 \times 10^{11}\,M_\odot$. We stress again that, in reality, it is unlikely that dissociated structures less massive than a galaxy cluster (i.e. $\ll 10^{14}\,M_\odot$) would be observed since the halo gas component becomes gradually less significant and harder to map at lower halo masses. However, in the scale-invariant non-radiative simulations, sub-cluster-scale haloes are geometrically similar to cluster-scale ones, thus a larger sample size can be utilized.

The probability distribution of $S$ in L210N1024NR is shown in Fig. 5. This distribution has a mean value of $\bar{S} \approx 0.068$ and 95 per cent of haloes are contained within $-0.03 \leq S \leq 0.19$. Only about 10 per cent of the haloes are negatively dissociated ($S < 0$). This skewness towards positive dissociation indices is also emphasized by the shaded region in Fig. 5, which depicts all negatively dissociated haloes and their reflection about $S = 0$, contrasting them with the positive dissociation indices. Qualitatively, this skewness is expected, due to the pressure forces that stop the gas during halo–halo collisions, reducing $q_{\text{gas}}$ relative to $q_{\text{DM}}$ (see equation 2).

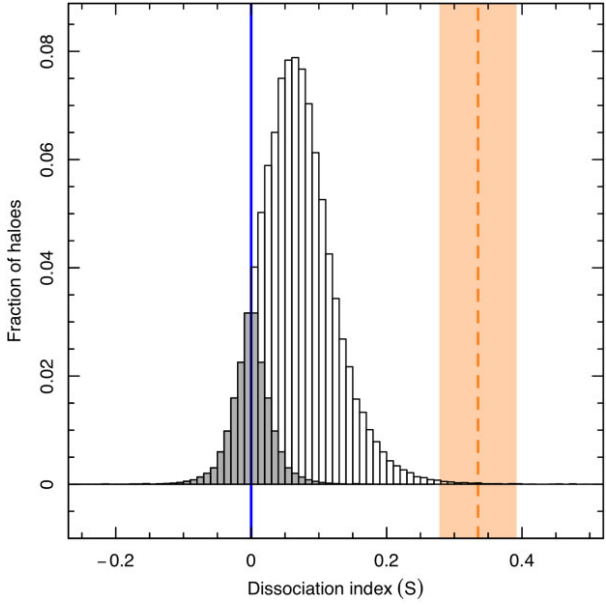

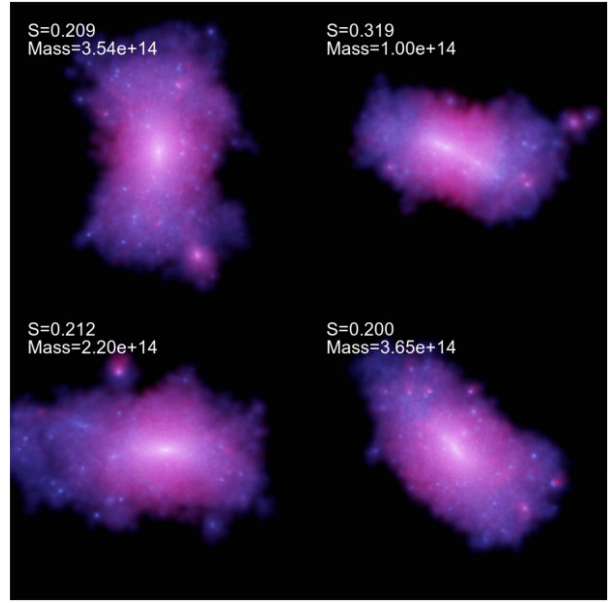

**Figure 5.** Distribution of dissociation indices for all first-generation haloes in SURFS L210N1024NR at redshift $z = 0$ that are well resolved ($\geq 7.5 \times 10^{11}$ M$_\odot$, corresponding to $\gtrsim 1000$ particles per species) with respect to $S$. $S = 0$ is indicated by a vertical blue line, the distribution of negatively dissociated haloes is shaded (grey) and reflected about $S = 0$. The estimated dissociation of the Bullet cluster $S_{\text{Bullet}} = 0.335 \pm 0.06$ is indicated via the orange dashed line and shaded region.

**Figure 6.** Four examples from SURFS L210N1024NR of cluster-scale haloes ($\geq 10^{14}$ M$_\odot$, corresponding to $\gtrsim 270\,000$ particles depending on the ratio of gas to DM) that are strongly dissociated ($S \geq 0.2$) at $z = 0$. As per convention, gas is shown in red and dark matter in blue. The column density is represented by the intensity of these colours. The dissociation index and mass of each example is given in the top left of the respective panel, with mass in units of M$_\odot$.

About 2 per cent of the FOF haloes are strongly dissociated, here defined as $S \geq 0.2$. This somewhat arbitrary threshold roughly delineates the structures that are obviously dissociated to the naked eye (if seen face-on). Fig. 6 illustrates what such strongly dissociated systems look like in L210N1024NR. Only $\sim 0.3$ per cent of the FOF haloes in this simulation have a dissociation index comparable to or larger than our estimate of $S_{\text{Bullet}} = 0.335 \pm 0.06$ for the Bullet cluster.

To develop a more complete picture of dissociation within L210N1024NR it is crucial to understand how the level of dissociation relates to halo mass. This is depicted via the $z = 0$ mass functions shown in Fig. 7. A halo mass function (HMF) is constructed using all haloes, binned by $S$, over the cosmological volume of $(210 \text{ cMpc } h^{-1})^3$. The HMF for dissociation indices $0 < S \leq 0.1$ (cyan) and $0.1 < S \leq 0.2$ (light blue) retain the profile of the total mass function (black dashed curve) as expected given these haloes fall within the inner 95 per cent of haloes distributed by $S$ (as shown in Fig. 5).

Fig. 7 shows that the partial HMFs for haloes within $-0.1 < S \leq 0.4$ (yellow, cyan, light blue, dark blue) are approximately parallel to the global HMF, implying that the dissociation distribution is mass-independent. This is indeed expected from the aforementioned scale-invariance of non-radiative cosmological simulations.

Only the most extreme dissociation indices (red, orange, purple), mainly found at mass below our sample cut (left vertical line in Fig. 7), seem to break this scale-invariance, as their HMF is steeper than the global HMF. As mentioned before, this is predominately a numerical effect of Poisson noise.

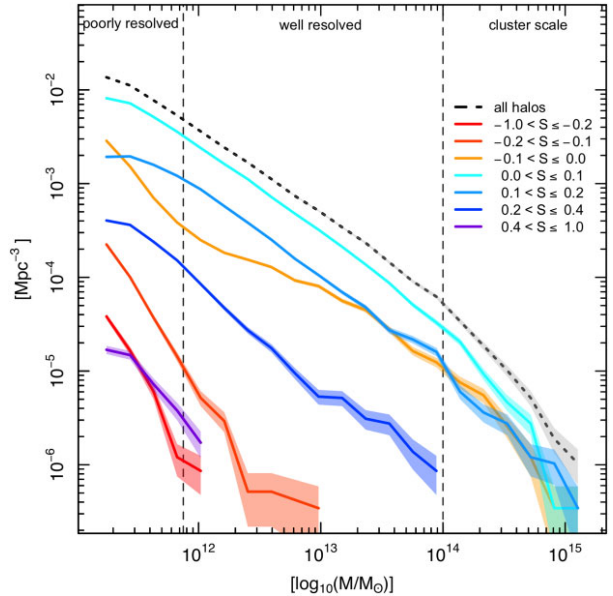

**Figure 7.** The halo mass functions (HMF) determined from the SURFS L120N1024NR simulation, using all haloes $> 10^{11}$ M$_\odot$ at redshift $z = 0$ within the cosmological volume of $(210 \, h^{-1} \text{cMpc})^3$, binned by the dissociation index. The global HMF is shown through the dashed curve and each partial HMF binned by $S$ is signified by a coloured solid line. For illustrative purposes all haloes $\geq 10^{11}$ M$_\odot$ are included. We delineate between poorly resolved haloes ($< 7.5 \times 10^{11}$ M$_\odot$), well resolved haloes ($\geq 7.5 \times 10^{11}$ M$_\odot$), and cluster-scale haloes ($\geq 10^{14}$ M$_\odot$) via the dashed vertical lines at the respective mass cuts.

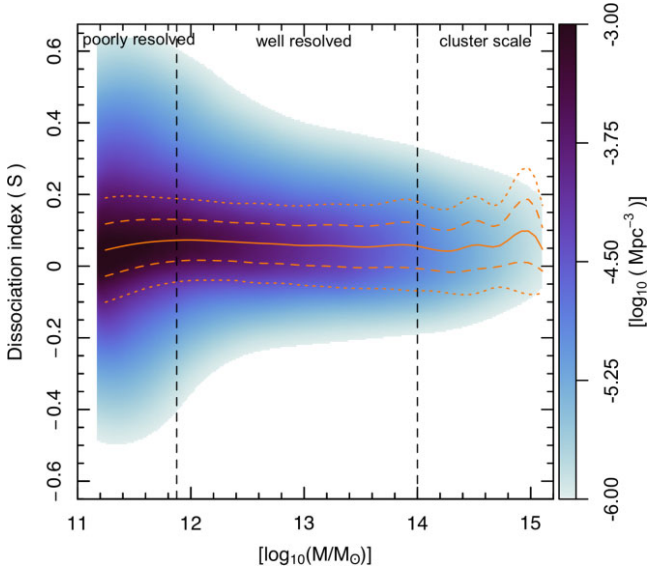

**Figure 8.** The 2D distribution of haloes in SURFS L210N1024NR at redshift $z = 0$ on the mass-dissociation index plane. The number density of haloes per mass and dissociation index is indicated by colour. The expected level of dissociation, $\bar{S}$ (orange curve) is overlaid with 68 per cent (orange dashed lines) and 95 per cent (orange dotted lines) confidence intervals. For illustrative purposes, as shown in Fig. 7, we include all haloes $\geq 10^{11}$ M$_\odot$. The mass cuts that delineate poorly resolved ($< 7.5 \times 10^{11}$ M$_\odot$), well resolved haloes ($\geq 7.5 \times 10^{11}$ M$_\odot$) and cluster-scale haloes ($\geq 10^{14}$ M$_\odot$) are shown via the dashed vertical lines.

Fig. 8 depicts the 2D distribution of haloes in the mass-dissociation index plane, which is the product of the 1D distributions given in Figs 5 and 7. The orange lines, showing the mean and scatter of $S$ as a function of mass, re-emphasize the notion that the distribution of $S$ is mass-independent for haloes above our mass cut (left vertical dashed line).

Given the number densities within the mass-dissociation index plane of Fig. 8, we now ask ourselves whether known dissociated clusters, such as the Bullet cluster, are expected in L210N1024NR? The Bullet Cluster has a mass of $\sim 1.5 \times 10^{15}$ M$_\odot$ (e.g. Clowe et al. 2004; Bradač et al. 2006; Clowe et al. 2006) and we estimated a dissociation index of $S_{\text{Bullet}} \sim 0.335 \pm 0.06$. As seen in Fig. 8 there are no Bullet cluster analogues (structures of approximate mass and dissociation) in L210N1024NR, this is unsurprising as a simulation requires a significantly larger volume than that of L210N1024NR ($(210c\text{Mpc } h^{-1})^3$) to expect such an object (e.g. Lee & Komatsu 2010; Thompson & Nagamine 2012; Bouillot et al. 2015; Kraljic & Sarkar 2015; Thompson et al. 2015). From the distribution presented in Fig. 8, it is trivial to estimate the required cosmological volume (the effective volume, $V_{\text{eff}}$) to expect structures of a given mass and dissociation index. By separating the 2D distribution on the mass-dissociation index planes into the component 1D distributions of mass and dissociation the effective volume is computed as

$$V_{\text{eff}}^{-1} = \int \int dS\, dM \phi(S, M)$$
$$= \int_{S_a}^{S_b} dS \phi_S(S) \int_{M_a}^{M_b} dM \phi_M(M) , \quad (12)$$

where $\phi_S(S)$ is the number density function associated with $S$ and $\phi_M(M)$ is the mass function presented in Fig. 7. Assuming a probable range of $S = 0.335 \pm 0.06$ and $1 < M < 2 \times 10^{15}$ M$_\odot$ we estimate a number density $\sim 4.92 \times 10^{-10}$ Mpc$^{-3}$ or that an effective volume of $\sim 2.03$ Gpc$^3$ would be required to observe a single Bullet-like cluster. This result is inline with the number density estimate of the order of $\sim 10^{-10}$ Mpc$^{-3}$ by Thompson et al. (2015), which improves on previous estimates (e.g. Lee & Komatsu 2010; Thompson & Nagamine 2012; Bouillot et al. 2015) due to more sophisticated halo finding methods (e.g. Behroozi, Wechsler & Wu 2013). Conversely, it was estimated by Kraljic & Sarkar (2015) (utilizing the same halo finder as Thompson et al. 2015) that given an effective volume of $\sim 14.6$ Gpc$^3$, no Bullet cluster analogues are expected, however as indicated by a pairwise velocity distribution it would be expected that present binary halo–halo orbits have the potential to form a Bullet-like object.

The observed scale-free relationship between the distribution of $S$ and halo mass depicted in Fig. 8 invites the question; how does this distribution come about and what scenarios within the assembly histories imposed by a $\Lambda$CDM universe specifically allow for strongly dissociated clusters? As dissociation occurs to some level in all collisions only the most recent collisions would be required to understand how an observed structure became dissociated and depict the distributions shown in Fig. 8. Idealized $N$-body + Hydrodynamic simulations are well suited for this exploration.

## 4 DISSOCIATION IN IDEALIZED BINARY COLLISIONS

Binary collisions of comparably massive haloes are the most common path to form dissociated structures in a hierarchical universe, as demonstrated, for example, by the highly dissociated simulated structures in Fig. 6. Empirically, this statement is confirmed by the finding that most strongly dissociated galaxy clusters show strong signs of past or ongoing collisions of two progenitor clusters. This is not to say that dissociated clusters are formed exclusively from binary systems. More complex 3-body (or more) systems have been observed with some level of dissociation such as MACS J1149.5+2223 (Golovich et al. 2016). However, these systems are comparatively rare.

To explain the distribution of $S$ found in the Universe and cosmological simulations, we can thus focus on binary halo–halo collisions. We do so via idealized $N$-body + SPH simulations of two colliding haloes, run with GADGET-3 (a modified version of GADGET-2; Springel 2005). All these controlled simulations are performed in 'Newtonian' mode, i.e. without cosmic expansion, and vacuum boundary conditions. Each of the colliding haloes is made of DM and an ideal gas without radiative cooling.

By studying the dissociation arising in a whole range of different binary halo collisions, we can then assess to what degree such binary collisions can indeed explain the distribution of dissociation indices found in our cosmological hydrodynamic simulation (Section 3).

### 4.1 Initial conditions

To initialize the controlled simulations of halo–halo collisions, we place two spherically symmetric gaseous progenitor haloes on a Keplerian orbit. If the haloes were point masses, their orbits would remain Keplerian, but tidal forces acting upon extended haloes will cause them to deform and change their orbits.

In the following, we first describe how the individual spherical haloes are initialized (Section 4.1.1) and then elaborate on how two such haloes are placed on a custom Keplerian orbit (4.1.2). A condensed reminder of how such orbits can be parametrized is provided in Appendix B.

*4.1.1 Extended halo progenitors*

We first construct a single 'base' halo to use as a template for the two orbiting haloes, by doing so we reduce any additional degrees of freedom surrounding the internal structure of either progenitor. The base halo is first initialized as a non-rotating spherical halo of mass $10^{12}$ M$_\odot$ containing $10^6$ particles sampled from an NFW density profile (Navarro, Frenk & White 1996), with a concentration parameter of $c = 10$, truncated to the radius $r_{200}$, defined such that the enclosed halo mass has a mean density 200 times above the present-day density of the Universe. Note that these non-radiative systems are inherently scale-free, so we could have chosen any mass and mean density, but we here picked Milky Way-like values to fix the ideas.

The $10^6$ particles are split into equal numbers of gas and DM particles with relative particle masses corresponding to a baryon mass fraction of 0.17.

We assign isotropic random velocities to the particles, drawn from a Gaussian distribution matched to the circular velocity. This method is selected for simplicity despite the limitations it places on the stability of the density profile at small radii (Kazantzidis, Magorrian & Moore 2004), however within the scope of the idealized simulations the resulting divergence from the NFW profile has a negligible impact on the estimates of $S$ (see Section C1).

The base halo was then evolved for 12 Gyr in a vacuum to allow for the halo to settle into a stable state with constant gas and DM density profiles that slightly differ from the initial setup. The evolved halo is then truncated back to a radius of $2r_{200}$. The resulting halo is then used as the base halo for the controlled halo–halo collisions.

To vary the mass ratio between the two colliding haloes ($\beta = M_2/M_1 \in [0, 1]$), we sample the second halo by sampling a fraction, $\beta$, of the particles of the base halo. The standard gravitational scaling relations imply that the stability is maintained by scaling the positions and velocities of the selected particles by a factor of $\beta^{\frac{1}{3}}$. Likewise, the internal energy of the selected gas particles needs to be scaled by a factor of $\beta^{\frac{2}{3}}$.

The smallest considered mass ratio in this paper is $\beta = 0.1$, meaning that the smaller halo only has $10^5$ particles. As demonstrated in Appendix C3, this is still more than sufficient for converged dissociation indices.

*4.1.2 Characterizing binary orbits of halo–halo collisions*

As explained in Appendix B, two parameters suffice to describe any Keplerian orbit – and thus orbits of two interacting bodies – if the overall orientation is irrelevant. For most of this work we choose these parameters as the pericentre distance $r_p$ and the eccentricity $e$. The latter captures the full range from circular ($e = 0$), to elliptical ($0 < e < 1$), parabolic ($e = 1$), hyperbolic ($1 < e < \infty$), and rectilinear ($e = \infty$) orbits. To exploit the scale-free nature of our simulations it makes sense to normalize $r_p$ to $\hat{r}_p \equiv r_p/r_{\text{hm},1}$, where $r_{\text{hm},1}$ is the half-mass radius of the more massive halo.

The first body of the two-body system is taken to be the more massive one (if any), such that the mass ratio $\beta = M_2/M_1$ is bound to $\beta \in [0, 1]$. Any idealized configuration of two orbiting point masses is then fully specified by the three dimensionless parameters $(e, \hat{r}_p, \beta)$. We now substitute each of these point-masses with our extended base haloes, such that their centre-of-mass positions and velocities are equal to those of the imaginary point-masses.

Relative to the centre-of-mass position and velocity of the combined two-halo system, the position vectors, $\boldsymbol{r}_1$ and $\boldsymbol{r}_2$, and velocity vectors, $\boldsymbol{v}_1$ and $\boldsymbol{v}_2$, of the two haloes are related via

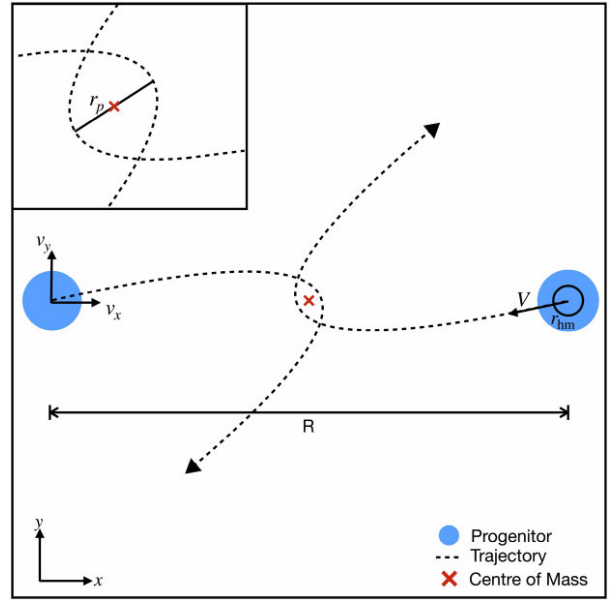

**Figure 9.** An illustration of the initial conditions of two progenitors positioned along the x-axis, separated by a distance $R$ for a parabolic orbit ($e = 1$) with an arbitrary $\hat{r}_p$ value.

$$\boldsymbol{r}_1 = -\beta \boldsymbol{r}_2, \tag{13}$$

$$\boldsymbol{v}_1 = -\beta \boldsymbol{v}_2. \tag{14}$$

The initial separation between the two haloes $R = |\boldsymbol{r}_1 - \boldsymbol{r}_2|$ can be chosen freely between the pericentre distance and apocentre distance. For open orbits ($e \geq 1$), the apocentre distance is technically infinite.

Without loss of generality, we define the $(x, y)$-plane of a Cartesian coordinate system as the orbital plane, and set up the initial haloes along the $x$-axis. Such an initial setup is sketched in Fig. 9. Given a choice of $(e, \hat{r}_p, \beta)$ and an initial distance $R$, we can then solve for the initial position and velocity components using standard Keplerian mechanics. Relative to the combined centre of mass, the initial position of the first halo is

$$\boldsymbol{r}_1 = \{-R\beta/(1+\beta), 0, 0\} \tag{15}$$

and its velocity is

$$\boldsymbol{v}_1 = \left\{ \sigma \sqrt{(\beta^{-1}+1)^{-1} \left( \frac{\beta(e-1)}{\hat{r}_p} + \frac{2\beta}{\alpha} \right) - \frac{\beta^2 \hat{r}_p (e+1)}{\alpha^2 (\beta+1)}}, \right.$$
$$\left. \sqrt{\frac{\sigma^2 \beta^2 \hat{r}_p (e+1)}{\alpha^2 (\beta+1)}}, 0 \right\}, \tag{16}$$

where $\sigma^2 \equiv GM_1/r_{\text{hm},1}$ is a scale velocity and $\alpha \equiv R/r_{\text{hm},1}$ is the normalized initial separation. It is straightforward to show that, for elliptical orbits,

$$\alpha \leq \hat{r}_p (1+e)/(1-e), \tag{17}$$

due to the requirement that the initial separation cannot exceed the apocentre distance. Solutions to equation (16) for an elliptical orbit ($e < 1$), are non-physical if $\alpha$ exceeds this range.

The positions and velocities of the other halo are then obtained via equations (13) and (14).

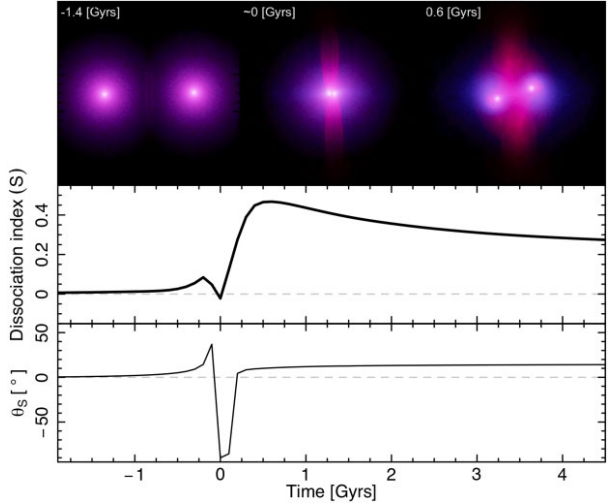

**Figure 10.** The results of an idealized simulation of a collision between two progenitors characterized by $(e, \hat{r}_p, \beta) = (1.1, 0.1, 1)$. The origin of time has been set to when the minimum root mean square (rms) distance between the DM haloes occurs. Top: three panels that each illustrate a specific snapshot output of the simulation, depicting the collision onset (left-hand side), along with the points at which $S$ is maximized (centre) and minimized (right-hand side). The column density of the DM (blue) and ICM (red) is indicated by colour intensity. Middle: depicts the evolution of the dissociation index as a function of time. Bottom: the orientation of the system, given by the angular separation between the major quadrupole axis and the positive x-axis, $\theta_S$.

### 4.2 Example of an idealized dissociative collision

Fig. 10 illustrates an isolated strongly dissociative collision between two equal-mass haloes ($\beta = 1$) with an orbit characterized by $e = 1.1$ and $\hat{r}_p = 0.1$. The figure shows both $S$ and the orientation of the major quadrupole axis (the axis with the greatest contribution to the quadrupole moment of the system) with respect to the positive *x*-axis ($\theta_S$) as a function of time. In this example, we have set the origin of time to be at the minimum root mean square (rms) distance between the DM haloes of the progenitors. The profile of $S$ and $\theta_S$ are used to subdivide the collision into a four notable stages: pre-collision ($t < -1.5$ Gyr), pericentric passage ($-1.5 < t \lesssim 0$ Gyr), the dissociative phase ($0 < t \lesssim 0.6$ Gyr) and post-collision phase ($0.6 < t$ Gyr).

Within the example of Fig. 10, the pre-collision phase is all points in the orbit prior to the onset of a collision where no discernible change in $S$ occurs. The first significant asymmetries between the spatial distributions of DM and gas arise during the pericentric passage. As the two DM haloes approach one another, their spatial distributions increasingly overlap, becoming more centrally concentrated and consequently the quadrupole amplitude ($q_{DM}$) decreases. The minimum $q_{DM}$ occurs at the point when the distance between both DM haloes geometric centres is minimized, as indicated by the rms distance. Hence the minimum value of $q_{DM}$ is largely determined by $\hat{r}_p$; if $\hat{r}_p = 0$ the DM spatial distribution momentarily approximates a monopole ($q_{DM} = 0$), whilst if $\hat{r}_p > 0$ some reflection of a quadrupole always exist within the DM distribution and $q_{DM} > 0$ at all times. At the initial stages of the collision prominent shocks form in the ICM. In the example of Fig. 10, due to the symmetry of the system (as $\beta = 1$), we observe a double shock front form perpendicular to the direction of the collision (such shocks have been noted in similar circumstances e.g. Poole et al. 2006; ZuHone 2011; Machado et al. 2015; Moura, Machado & Monteiro-Oliveira 2021). As the haloes continue to collide, the ICM of each progenitor is compressed along the direction of the collision and pushed outwards from the point of collision due to dynamic pressure.

Here the compressed gas forms a 'disc-like' distribution (of course, the exact morphology and behaviour of the compressed and shocked gas changes depending on the initial conditions of the orbit), which has an increasing quadrupole amplitude ($q_{gas}$) as more of the colliding ICM is compressed and extends further outwards about the axis of the collision. The increasing quadrupole moment of the compressed ICM relative to the decreasing DM quadrupole moment results in a decreasing $S$. Additionally as $q_{gas}$ momentarily becomes dominant, $q_{gas} > q_{DM}$, the major axis becomes orientated along the direction in which the compressed ICM is most extended.

As the geometric centres of the two DM haloes cross one another ($t = 0$ Gyr) $q_{DM}$ begins to increase relative to $q_{gas}$, as the DM spatial distribution becomes more quadrupole like. The DM haloes then move ahead of the gas (as shown in the top row of Fig. 10 at $t \approx 0.6$ Gyr) which is slowed by dynamic pressure due to the consequential exchange of momentum and altered spatial distribution. Resulting in a rapid increase of $S$. It is in this dissociative phase that a variety of transient structures can form within the ICM. One example we consistently observed in our strongly dissociative collisions (which will be introduced in Section 4.3) when $\beta \geq 1/3$, was the formation of a gaseous bridge between the cores of each progenitor, akin to those discussed in greater detail by Poole et al. (2006).

Notably in Fig. 10, following the rapid increase during the dissociative phase, $S$ 'relaxes' as it asymptotically approaches a final value ($S_{final}$). Provided the DM haloes have a positive orbital energy, they will continue to separate further and the ICM will continue to expand, developing an increasing quadrupole moment which combined with the $\bar{r}$ normalization term slowly reduces $S$ (there are many other possibilities, which are dependent on the exact orbital geometry). In a scenario where the DM haloes have a negative orbital energy following the dissociative phase, they will re-collapse on to one another and undergo a series of decreasingly energetic collisions accompanied by lesser dissociative phases until a merger is complete. Additional pericentric passages erase or obscure any previously attained dissociation as the DM haloes continually 'mix' the centralized ICM.

### 4.3 Dissociation as a function of collisional parameters

To build on the example given in Section 4.2, we now must ask how would altering the initial conditions of an orbit as defined by the parameters $(e, \hat{r}_p, \beta)$ affect the resultant dissociation.

We can assume there is some unique relationship between the dimension-less collision parameters $(e, \hat{r}_p, \beta)$ and the resulting dissociation $S$ given the scale-invariant nature of Newtonian gravity and idealized gas dynamics of the idealized simulations. We explore this relationship by sampling the space defined by $(e, \hat{r}_p, \beta)$ and initializing an idealized simulation for each sampled location.

#### 4.3.1 Sampling the parameter space

We constrain the space of $(e, \hat{r}_p, \beta)$ by selecting just three progenitor mass ratios to explore; $\beta = 1, 1/3,$ and $1/10$. The remaining two parameters are sampled in the domain bound by $0 < e \leq 4$ and $0 \leq \hat{r}_p \leq 4$ for each value of $\beta$ as shown in Fig. 11. It has been shown that a region of $0 < e \leq 2$ is sufficient when describing orbits of merging haloes (e.g. Khochfar & Burkert 2006; Poulton 2019).

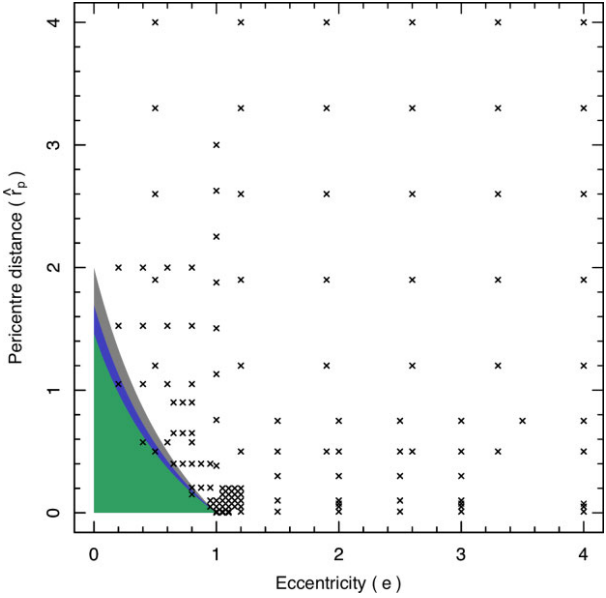

**Figure 11.** The parameter space defined by $(e, \hat{r}_p)$, with all constraints as outlined in Section 4.3.1. The initial conditions for all idealized simulations at $\beta = 1, 1/3,$ and $1/10$ are indicated by an 'x'. The region of elliptical orbits restricted by $r_a$ is shaded according to the mass ratio; $\beta = 1$ (grey), $\beta = 1/3$ (blue), and $\beta = 1/10$ (green).

However, we also consider scenarios in which strongly dissociated clusters are formed by highly hyperbolic orbits and thus extend this region to $0 < e \leq 4$. The interval $0 \leq \hat{r}_p \leq 4$ covers the vast majority of collisions, numerically it can be seen that for $\hat{r}_p > 4$ no significant dissociation occurs in our idealized simulations (see Sections 4.3.2–4.3.3). The parameter space $(e, \hat{r}_p)$ is additionally constrained by a minimum apocentric distance, $r_a \geq r_{\text{hm}, 1} + r_{\text{hm}, 2}$ and is implemented at each mass ratio as shown by the shaded regions of Fig. 11. This criterion is imposed to ensure both progenitors are initially separated so the orbit does not begin with either halo already colliding with the other.

We prioritize sampling initial conditions that correspond to common collisions in $\Lambda$CDM (see Section 5) or produce highly dissociated clusters, hence a preference for hyperbolic orbits with $\hat{r}_p < 1$ (as will be shown in Section 4.3.3). The initial positions of each progenitor are determined via equations (13)–(16), where $\alpha$ for elliptical orbits is computed by equation (17) (with equality sign). For open orbits ($e \geq 1$), we choose $\alpha = 15$ to ensure a sufficient initial separation for pressure and tidal forces between the two haloes to be negligible.

### 4.3.2 Dissociation as a function of time

In each of the idealized simulations, $S$ is calculated in intervals of 0.1 Gyr. By comparing the profiles of $S(t)$ for all orbits, over the initial pericentric passage and dissociative phase a 'characteristic' profile can be recognized which qualitatively follows the scenario in Fig. 10. We identify two key features in this profile: (1) during the pericentric passage $S(t)$ decreases or stagnates as the ICM of each progenitor is compressed and pushed outwards. (2) The DM cores of each progenitor cross over one another and continue onwards ahead of the collisional ICM, resulting in $S(t)$ rapidly increasing. The evolution of $S(t)$ following the peak diverges depending on the orbital parameters.

Fig. 12 displays $S(t)$ for all strongly dissociated collisions, which exemplify this characteristic profile (see Section D for additional results). Interestingly, in the isolated collisions of our simulations, dissociated morphologies are retained over a significant period of time ($t \geq 3$ Gyr) as indicated in Fig. 12. Following an increase in the dissociation, $S(t)$ 'relaxes' by asymptotically approaching a final value such that $S_{\text{max}} \approx S_{\text{final}}$, unless the DM haloes turn around and collide again thus disrupting this asymptotic relaxation.

Within the sample of both parabolic and hyperbolic orbits, we observe a sub-population of orbits with $\hat{r}_p \lesssim 0.1$ where an initial decrease in $S$ during the pericentric passage is not observed. However, using smaller intervals of time, $t = 0.05$ and $t = 0.025$ Gyr this decrease in $S$ during the pericentric passage is recovered (see Fig. D2). This sub-population highlights one aspect of the relationship between $S$ and the parameter space $(e, \hat{r}_p, \beta)$, when the orbital velocity (computed from $e$, $\hat{r}_p$, and $\beta$ as shown by equation (16)) increases, the time from $S_{\text{min}}$ to $S_{\text{max}}$ is reduced whilst the magnitude of both $S_{\text{min}}$ and $S_{\text{max}}$ increases. A small note needs to be made for the orbits seen in Fig. 12 characterized by $\beta = 1/10$ that start with a negative dissociation index ($S \approx -0.045$), which has no significant impact on the results of this paper. This is due to Poisson noise and the amplification of small asymmetries between gas and DM within the base halo when it is scaled to smaller mass ratios (see Section C2).

### 4.3.3 The potential dissociation of different collisions

To simplify the discussion of the relationship between $S$ and the collisional parameters $(e, \hat{r}_p, \beta)$, we focus on the maximum dissociation that arises shortly after the first pericentric passage. Which as shown in Fig. 12 is close to the long-term asymptotic value for hyperbolic orbits ($S_{\text{max}} \approx S_{\text{final}}$) provided no further collisions occur. Fig. 13 shows the relation between the collisional parameters and $S_{\text{max}}$. It is evident that $S$ is dependent on the progenitor mass ratio ($\beta$), orbital eccentricity ($e$), and pericentre distance ($\hat{r}_p$). Generally speaking, $S$ increases along the axis of orbital eccentricity, decreases at larger pericentre distances, and increases with $\beta$. Here we describe this relationship in a rather simplistic and qualitative manner using the relationship of each parameter to dynamic pressure, given it is the primary cause of the dissociation. Furthermore dynamic pressure is scale-invariant like the distribution of $S$ and our collisional parameters $(e, \hat{r}_p, \beta)$. Within an astrophysical context dynamical pressure is referred to as ram pressure, which is proportional to the fluid mediums density and the square velocity of the bulk flow normal to the surface; $P_{\text{ram}} \propto \rho v^2$ (Gunn & Gott 1972). Meaning for a given binary system a more energetic orbit implicates a greater orbital velocity, which results in a greater dynamic pressure exerted on the ICM of each progenitor and consequently leads to a greater level of dissociation. As seen in Fig. 13, $S$ is maximized as $e \to \infty$, $\hat{r}_p \to 0$, and $\beta \to 1$, all of which increase the orbital velocity of the progenitors in accordance with equation (16). Additionally, as the pericentre distance decreases a greater proportion of the ICMs volume is involved in the collision, therefore a greater volume of the ICM can be spatially decoupled from the DM, hence $S_{\text{max}}$ increases as a result. It can also be argued that at smaller pericentric distances the colliding haloes are falling through a denser medium than at larger pericentre distances, contributing in part to the relationship between $S$ and $\hat{r}_p$.

As shown by the red curves in Fig. 12 for a given orbit characterized by $(e, \hat{r}_p)$, the magnitude of dissociation is increased at all points as the progenitors become more equivalent in mass ($\beta \to$ 1). In addition to the increased orbital velocity, both the symmetry of the system and the relative volumes of the total ICM involved in the collision increase with $\beta$. The snapshots shown in Fig. 12, illustrate these ideas rather well. For the same orbit, at smaller values of $\beta$ the capacity of the second progenitor (less massive progenitor) to create asymmetries between the spatial distributions of the ICM and DM is increasingly limited. As the second progenitor affects a lesser proportion of the larger progenitors' ICM, the potential level of dissociation is reduced. Furthermore, the symmetry of the system increases with $\beta$, as such each species spatial distributions are more quadrupole-like, meaning the magnitude of $S$ at each point is scaled by some function of $\beta$.

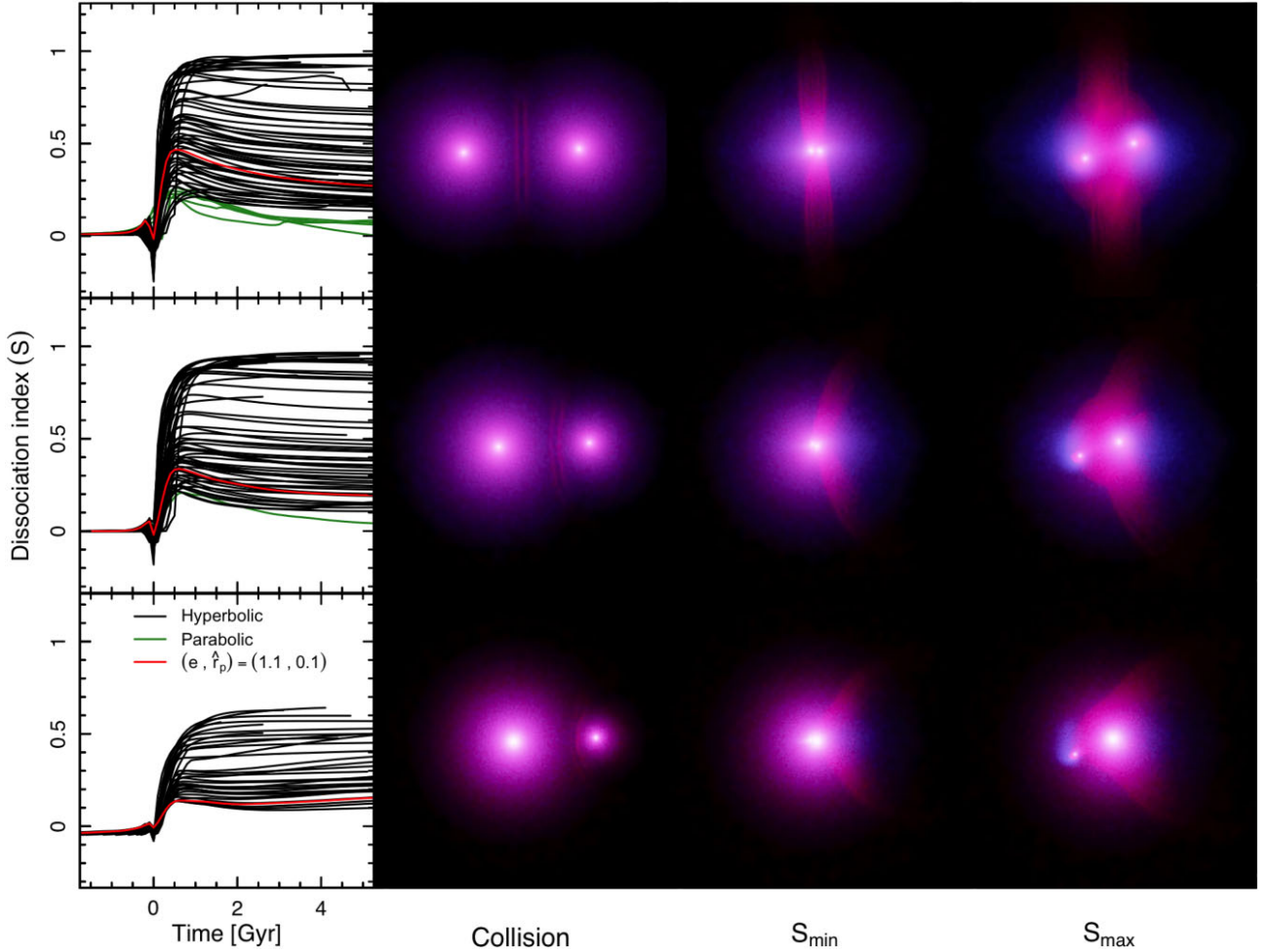

**Figure 12.** A comparison of the profile of $S(t)$ for all strongly dissociative clusters at each sampled mass ratio. The origin of time is set to the minimum rms distance between the two DM haloes. The key points within the 'characteristic' profile of $S(t)$ are shown as a 2D projection of the orbit characterized by $(e = 1.1, \hat{r}_p = 0.1)$ at each mass ratio ($\beta = 1, 1/3, 1/10$). All gas particles are coloured red, DM is coloured blue, the overlap of the two, therefore, appears purple and the brightness indicates the column density. The initial negative values of $S$ within the 1:10 simulations are a numerical artefact of Poisson noise (see Section C2 for a further discussion) and does not have a significant impact on $S$ measured within the collision.

Thus far we have kept the discussion of $S$ as a function of the collisional parameters $(e, \hat{r}_p, \beta)$ to a generalized qualitative level. A more complete discussion of $S$ concerning the idealized simulations demands a more detailed look at each result individually which is beyond the objectives of this paper. Rather we seek to understand the relationship between $S$ and the collisional parameters $(e, \hat{r}_p, \beta)$ with respect to the distribution of dissociated objects in a $\Lambda$CDM paradigm as shown in Section 3. We now turn our attention to this task.

## 5 BINARY COLLISIONS IN A COSMOLOGICAL CONTEXT

Having established a quantitative relation between the parameters of binary halo collisions and resulting dissociation indices, we can now address the important question of whether the binary collisions expected in a $\Lambda$CDM universe can account for the full dissociation statistics seen in such a universe. To do so, we first determine the statistics of binary halo collisions in a simulation of cosmic large-scale structure and then investigate if this distribution applied to our controlled collision simulations (Section 4) can qualitatively and quantitatively reproduce the dissociation statistics found in our hydrodynamic cosmological simulation (Section 3).

### 5.1 Statistics of binary halo collisions

In cosmological simulations, the distribution of the dimensionless collision parameters $(e, \hat{r}_p, \beta)$ can be determined using halo merger trees. In principle, we could use the merger trees of the SURFS simulation L210N1024NR introduced in Section 3.1. However, a pure DM simulation (without gas) suffices to extract the merger trees and we can therefore benefit from higher resolution SURFS runs that were performed with DM only. We use the SURFS run L210N1536 (properties in table 1 Elahi et al. 2018), which has the same characteristics and initial conditions as L210N1024NR. The only difference is that L210N1536 has no gas particles, but instead uses 3.375-times more DM particles ($1536^3$ in total).

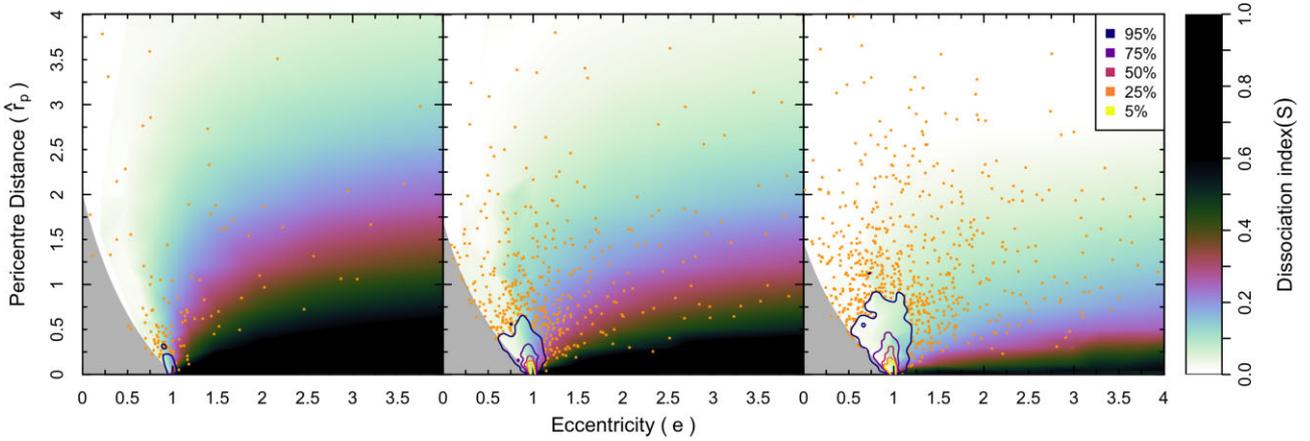

**Figure 13.** The distribution of the maximum dissociation indices ($S_{\rm max}$) achieved in binary orbits characterized by ($e, \hat{r}_p, \beta$) at each sampled point in the parameter space ($e, \hat{r}_p$) at mass ratios; $\beta = 1$ (left-hand panel), 1/3 (centre), and 1/10 (right-hand panel). In each panel the grey shaded region represents the space of elliptical orbits restricted by the minimum apocentric distance, $r_a$. The overlaid contours and individual points correspond to the work presented in Section 5. These contours outline the spatial distribution of both $e$ and $\hat{r}_p$ from the binary collisions of 1st generation haloes identified in SURFS L210N1536 halo merger trees at $z \leq 2$. Each of the overlaid contours are built from orbits which have a mass ratio within $\pm 0.05$ of $\beta$ in each panel and show the space containing 95 per cent, 75 per cent, 50 per cent, 25 per cent, and 5 per cent of identified orbits. The remaining individual orbits outside of the contours are each indicated as individual orange points.

Merger trees for L210N1536 readily exist (as described in Obreschkow et al. 2020) and were constructed using the TREEFROG code (Elahi et al. 2019b), applied to the VELOCIraptor halo catalogue (Elahi et al. 2019a, see also Section 3.1). The trees we are using here are clean trees that only connect first-generation haloes (without substructure) in a strictly hierarchical structure without missing snapshots. In other words, each halo has a unique descendant in the following simulation snapshot (as in the 'Dhalo' format, see Jiang et al. 2014).

From the halo merger trees, we select all the binary collisions between haloes at redshift $z \leq 2$ with a mass ratio $\beta \geq 0.01$ and where all progenitors are more massive than $10^{12}\,{\rm M}_\odot$. This mass cut ensures that all merger events are well resolved and robustly identified by the halo finder and tree builder.

For each merger event in this sample, we trace the two most massive progenitors back in time until their half-mass spheres first start to overlap, i.e. until the centre-to-centre distance between the two haloes first exceeds the sum of their half-mass radii, $r_{\rm hm,1} + r_{\rm hm,2}$. This tracking-back procedure is important as it allows us to determine the orbital elements of the merger before tidal forces can significantly affect the centre-of-mass positions and before the friends-of-friends algorithm is at risk to ascribe the particles of one halo to the other. We found that tracking the progenitor haloes even further back in time, to separations greater than the sum of their half mass radii, does not make much of a difference to our results (see Section E for further discussion). Finally, the orbital parameters, $e$ and $\hat{r}_p$ of the two main progenitors in the binary collision are estimated at the snapshots where these two progenitors first touch (in the above sense). Note that we rejected all merger events where the systems total mass changes by more than 10 per cent between neighbouring snapshots (demonstrated by Khochfar & Burkert 2006), anywhere between the snapshot of the touching progenitors and the snapshot of the merger. This allows us to exclude a small fraction of mainly low-mass systems in complex environments, where the friends-of-friends finder struggled to properly identify the haloes. This criterion is not critical, but it helps reduce the numerical uncertainty in the distributions of orbital parameters.

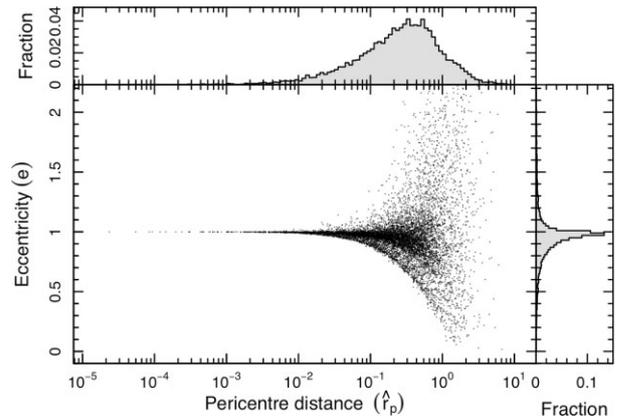

**Figure 14.** Distribution of orbital parameters of binary halo collisions in SURFS L210M1536 with mass ratios $\beta \geq 0.01$. This distribution is nearly mass independent.

Our distribution of the orbital parameters of binary mergers is presented in Fig. 14. The distributions of $e$ and $\hat{r}_p$ are self-similar for $\beta > 1/10$ and the distribution given by Khochfar & Burkert (2006) is recoverable from Fig. 14 if we only include halo mergers (as opposed to all identified collisions) and do not select only first-generation haloes. The dense population of approximately parabolic orbits $e = 1 \pm 0.13$ in the region $\hat{r}_p \gtrsim 0.2 \pm 0.1$ (see Section E2 for uncertainty on each estimate), along with a significant portion of the hyperbolic orbits are not found in the distribution of orbits that eventuate in a merger. These orbits either retain a positive orbital energy or have yet to completely merge, existing as long-lived subhaloes.

### 5.2 Dissociation from halo merger trees

We project the orbits shown in Fig. 14 that are within $\pm 0.05$ of $\beta = 1$, $1/3$, and $1/10$ on to our sampled parameter space in Fig. 13. Estimating the potential dissociation for each orbit identified in L2101536, giving an expected upper limit of $S$ in a $\Lambda$CDM universe.

Examining Fig. 13, it is apparent that the majority $\sim 87.8$ per cent of collisions between haloes do not have the potential to be strongly dissociative at any $\beta$ value sampled. This can be attributed to both the declining frequency of collisions with increasing $\beta$ and the rarity of hyperbolic orbits, which themselves are unfavourably distributed for maximizing dissociation. Generally for a given $\beta$, made apparent in Fig. 13 (and noted in Section 4.3.3), as $e \to \infty$ and $\hat{r}_p \to 0$, the orbital energy increases, thus $S$ is maximized. However, orbits within L210N1536 are unlikely to populate the high orbital energy, hyperbolic regions of the parameter space. Such hyperbolic orbits occur mainly as a result of rarefied three (or more) body interactions. By comparison, less energetic orbits are more probable, as shown by Fig. 14, which is in agreement with the expectation that most structures are weakly dissociated. We find no compelling evidence in L210N1536 for elliptical orbits to have the potential to be strongly dissociative. From the distribution projected on to Fig. 13, we find five elliptical orbits in the region $0.99 \lesssim e < 1$, $\hat{r}_p \le 0.37$, and $\beta \gtrsim 0.95$ with $S \sim 0.2$. If all orbits ($\beta > 0.01$) from L210N1536 are included we find numerous candidates with comparable parameters; $0.99 < e < 1$ and $\hat{r}_p < 0.4$. However, given the uncertainty of each estimated parameter ($\hat{r}_p \pm 0.1$ and $e \pm 0.13$, see Section E2) and the lower mass ratios it is unclear or tenuous at best if many elliptical orbits have the capacity be strongly dissociative. It is likely the majority of these orbits are parabolic or hyperbolic orbits that have had an underestimated eccentricity.

Furthermore, we identify no Bullet Cluster analogues in Fig. 13 as expected given the discussion in Section 3. However, $\sim 3.45$ per cent of orbits have the potential to be comparably dissociated as they lie within the estimated range of the Bullet cluster ($S_{\text{Bullet}} \sim 0.335 \pm 0.06$) and only $\sim 2.83$ per cent of orbits have the potential to be more dissociated than the Bullet Cluster. Emphasizing that even in the most optimistic scenarios, as discussed here, strongly dissociated structures are unique products of the most extreme and unlikely orbits. The scarcity of objects such as the Bullet cluster is to be expected (as discussed in Section 3).

### 5.3 Comparing measured and expected dissociation indices

The distribution of the collisional parameters $(e, \hat{r}_p, \beta)$ from L210N1536 and the unique relationship they have to $S$ can be used to describe the expected distribution of $S$ at redshift $z = 0$ ($S_{z=0}$), as shown in Fig. 15. The expected distribution of $S_{z=0}$ from SURFS L210N1024NR (as discussed in Section 3 and presented in Fig. 5) is centred on $S \sim 0.068$ with 95 per cent of haloes found in the region $-0.03 \lesssim S_{z=0} \lesssim 0.19$ and can be compared to the estimated distribution of $S$ from the binary orbits in L210N1536 using the results of the idealized simulations. The $S_{\text{max}}$ distribution shown in Fig. 15 is directly from the orbits projected on to the idealized parameter space as shown in Fig. 13. Qualitatively $S_{\text{max}}$ is the upper limit or distribution of the potential dissociation, where every halo is assumed to be observed when it is most dissociated. As seen in Fig. 15 by $S_{\text{max}}$ being more positively skewed than $S_{z=0}$. The coincident peaks of $S_{\text{max}}$ and $S_{z=0}$ indicate that the expected dissociation of haloes at redshift $z = 0$ is related to the distribution of orbits characterized by the parameters $(e, \hat{r}_p, \beta)$. Explicitly, the dissociation indices expected from the most frequently occurring halo–halo collisions are coincidental with those most frequently observed at $z = 0$.

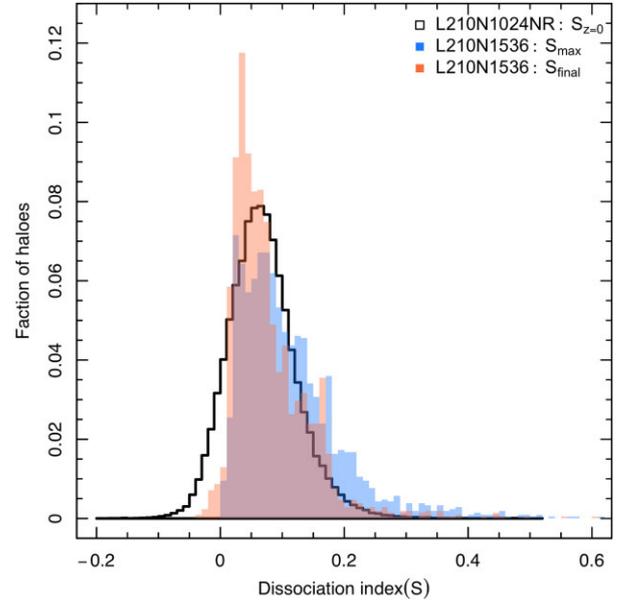

**Figure 15.** A comparison between the distribution of dissociated haloes at redshift $z = 0$, $S_{z=0}$, from the SURFS L210N1024NR simulation (as given in Section 3) to the expected distributions of $S$ from the distribution of parameters $(e, \hat{r}_p, \beta)$ in binary orbits from the SURFS L210N1536 simulation. The expected distributions use the estimated maximum ($S_{\text{max}}$) and final ($S_{\text{final}}$) dissociation indices for each collision charaterized by $(e, \hat{r}_p, \beta)$.

It cannot be expected that all haloes at any point in time are undergoing collisions or merging, let alone be at a point where the dissociation is maximized or minimized. The vast majority of haloes not undergoing collisions at $z = 0$ are expected to have a dissociation index value that is 'relaxing' towards or oscillating about $S \sim 0.068$ (as introduced with Figs 10 and 12). This timing aspect implies a reduction in the variance and skew of the distribution $S_{z=0}$ compared to $S_{\text{max}}$, as seen in Fig. 15. To account for the lack of haloes being at various stages of a collision in the $S_{\text{max}}$ distribution, we introduce an additional expected distribution from the orbital parameters $(e, \hat{r}_p, \beta)$ fitted to the final dissociation indices ($S_{\text{final}}$) from each idealized simulation as opposed to the maximum. Where $S_{\text{final}}$ represents the point at which the vast majority of collisions have been completed. There are ongoing mergers in the population of $S_{\text{final}}$, as they were not complete within a Hubble time from when they initially collided. The distribution of $S_{\text{final}}$ is less positively skewed and has a lesser variance than $S_{\text{max}}$, as expected given $|S_{\text{final}}| \lesssim |S_{\text{max}}|$.

As seen in Fig. 15 the region in which both the $S_{\text{max}}$ and $S_{\text{final}}$ distributions overlap outlines the shape of $S_{z=0}$ for positive dissociation indices. Suggesting that within a $\Lambda$CDM universe the distribution of $(e, \hat{r}_p, \beta)$ in halo–halo collisions and the resultant dissociation driven by ram-pressure is able to explain a significant proportion of the observed dissociation between DM and gas at $z = 0$. Intuitively, this relationship of the collisional parameters $(e, \hat{r}_p, \beta)$ with $S$ and the frequency with which orbits characterized by $(e, \hat{r}_p, \beta)$ occur would to some degree depict the distribution of $S$ at any point in time, given binary halo–halo collisions are the primary pathway for positive dissociation in structures.

The distribution of negatively dissociated haloes is not accounted for by either $S_{\max}$ or $S_{\text{final}}$. These negative values are likely from ongoing binary collisions observed during the pericentric passage (see Section 4.2) or from physical processes outside of ram pressure and additional scatter in $S$ from low mass structures. These scenarios are not fully realized in our idealized simulations which $S_{\max}$ and $S_{\text{final}}$ are derived from. However, it is most likely that at a given point in time a halo will be positively dissociated, given the small window in time that it can appear negatively dissociated during a collision. The most negatively dissociated morphologies occur during collisions with the greatest potential dissociation and in all idealized simulations we find no instances where $|S_{\min}| > |S_{\max}|$. The lower probability and smaller magnitudes of negatively dissociated haloes compared to positively dissociated haloes predict that the $S_{z=0}$ distribution is centred on a positive value, ($S \sim 0.068$), and exhibits a positive skew as seen in Fig. 15. Additionally it should be noted that within the population of haloes in $S_{z=0}$, there are highly complex structures formed through collisions of more than two haloes, which have multiple sub-clusters of DM or gas (e.g. Abell 520; Mahdavi et al. 2007; Jee et al. 2012) that are better described by multipole moments of a higher degree than the quadrupole moment. As such these structures are not accurately described by the dissociation index presented in Section 2, accounting for some of the discrepancies between $S_{z=0}$ and both $S_{\max}$ and $S_{\text{final}}$.

## 6 CONCLUSIONS

In this paper, we have introduced the dimensionless dissociation index $S$ to quantify the spatial quadrupole-like separation between two fluids. This index was applied to study the spatial separation between gas and DM in haloes. In order to compare numerical data to observations, we also introduced an observational estimator of $S$, applicable to 2D gas maps (e.g. from X-ray imaging) and DM density maps (e.g. from gravitational lensing). Our main findings are as follows:

(i) By applying the dissociation index to a hydrodynamic cosmological simulation, we found that most (∼90 per cent) of the haloes in a $\Lambda$CDM universe are positively dissociated, meaning that their DM is more 'stretched out' along a particular direction than the gas.

(ii) The distribution of dissociation indices is well approximated by a normal distribution of mean $\bar{S} = 0.068$ and standard deviation $\sigma_S = 0.05$. In scale-free non-radiative simulations, this distribution is invariant of the halo mass.

(iii) Highly dissociated galaxy clusters ($S > 0.2$) are rare. For example, haloes with a mass and dissociation index comparable to the Bullet cluster ($M = 1\text{–}2 \times 10^{15}\,\mathrm{M}_\odot$ and $S = 0.335 \pm 0.06$) are expected to be found once per comoving volume of $2\,\mathrm{Gpc}^3$ at a given time.

(iv) Through a suite of controlled simulations of idealized collisions between two gaseous haloes, we have analysed the evolution of the dissociation index as a function of time. It turned out that in significantly dissociative collisions, $S$ gets 'frozen', meaning that its long-term asymptotic value is close to its maximum value during the collision.

(v) Using these controlled simulations we have also determined the quantitative relation between the orbital parameters (eccentricity $e$, normalized periapsis distance $\hat{r}_p$) and the resulting dissociation index $S$, for different mass ratios $\beta$. This analysis reveals that highly dissociated structures require small impact parameters and energetic orbits (parabolic and hyperbolic).

(vi) Finally, by applying the relationship between ($e$, $\hat{r}_p$, $\beta$) and $S$, determined from the controlled simulations, to the distribution of ($e$, $\hat{r}_p$, $\beta$) in collisions in a simulated $\Lambda$CDM universe, we are able to reproduce the distribution of $S$ of such a universe to a remarkable degree. This suggests that the dissociation between gas and DM in a $\Lambda$CDM cosmology is largely explainable via ram-pressure driven dissociation in binary halo–halo mergers.

Overall the dissociation index $S$ introduced in this paper appears to be a useful measure of the dissociation between DM and gas, and thus a promising tool to compare observations of dissociated clusters to predictions from cosmological simulations. We have shown how $S$ can be used as a statistic to categorize and study real cluster data via the Bullet cluster, from which a natural next step is to then apply this statistic to a current, larger catalogue of real clusters (dissociated or otherwise) that have X-ray and gravitational lensing data readily available. Additionally extending the application of $S$ to ongoing/forthcoming X-ray surveys and lensing campaigns, as well as to cosmological simulations with full baryon physics, such as the EAGLE (Schaye et al. 2015) and IllustrisTNG (Pillepich et al. 2018) simulations, are promising avenues for future research.

## ACKNOWLEDGEMENTS

We thank the anonymous referee for a very constructive report. DO is a recipient of an Australian Research Council Future Fellowship (FT190100083) funded by the Australian Government. WM acknowledges ICRAR-UWA for access to the suite of SURFS simulations and the Hyades computing cluster, which was utilized throughout this research. We acknowledge that the SURFS runs were conducted on Magnus at the Pawsey Supercomputing centre, which is supported by the Australian Government and the Government of Western Australia. WM thanks Chris Power and Claudia Claudia Lagos for their invaluable insights that aided in establishing the idealized simulations and analysing halo merger trees, respectively. We thank Mark Boulton for his assistance with Hyades and the software packages required to conduct the analysis presented in this paper. We have used R (versions 3.6.3 through 4.2.0, R Core Team 2022) in our analysis and acknowledge the use of MAGICAXIS (Robotham 2022) and GADGETRY (Obreschkow 2022) to create the figures shown.

## DATA AVAILABILITY

The suite of idealized $N$-body + SPH simulations presented within this paper is available on request. Additionally, the SURFS simulations used are available on request (icrar-surfs@icrar.org).

## REFERENCES

Angulo R. E., Hahn O., 2022, Living Rev. Comput. Astrophys., 8, 1
Behroozi P. S., Wechsler R. H., Wu H.-Y., 2013, ApJ, 762, 109
Binney J., Tremaine S., 1987, Galactic Dynamics. Princeton Univ. Press, Princeton
Bouillot V. R., Alimi J.-M., Corasaniti P.-S., Rasera Y., 2015, MNRAS, 450, 145
Bradač M. et al., 2006, ApJ, 652, 937
Bradač M., Allen S. W., Treu T., Ebeling H., Massey R., Morris R. G., von der Linden A., Applegate D., 2008, ApJ, 687, 959
Clowe D., Gonzalez A., Markevitch M., 2004, ApJ, 604, 596
Clowe D., Bradac M., Gonzalez A. H., Markevitch M., Randall S. W., Jones C., Zaritsky D., 2006, ApJ, 648, L109

Clowe D., Markevitch M., Bradač M., Gonzalez A. H., Chung S. M., Massey R., Zaritsky D., 2012, ApJ, 758, 128
Cohn J. D., White M., 2005, Astropart. Phys., 24, 316
Correa C. A., Wyithe J. S. B., Schaye J., Duffy A. R., 2015, MNRAS, 450, 1514
Crocce M., Pueblas S., Scoccimarro R., 2012, Astrophysics Source Code Library, record ascl:1201.005
Davis M., Efstathiou G., Frenk C. S., White S. D. M., 1985, ApJ, 292, 371
Dawson W. A., 2013, ApJ, 772, 131
Dawson W. A. et al., 2012, ApJ, 747, L42
Diego J. M. et al., 2020, ApJ, 904, 106
Drakos N. E., Taylor J. E., Berrouet A., Robotham A. S. G., Power C., 2019a, MNRAS, 487, 993
Drakos N. E., Taylor J. E., Berrouet A., Robotham A. S. G., Power C., 2019b, MNRAS, 487, 1008
Elahi P. J., Welker C., Power C., Lagos C. d. P., Robotham A. S. G., Cañas R., Poulton R., 2018, MNRAS, 475, 5338
Elahi P. J., Cañas R., Poulton R. J. J., Tobar R. J., Willis J. S., Lagos C. d. P., Power C., Robotham A. S. G., 2019a, Publ. Astron. Soc. Austr., 36, e021
Elahi P. J., Poulton R. J. J., Tobar R. J., Cañas R., Lagos C. d. P., Power C., Robotham A. S. G., 2019b, Publ. Astron. Soc. Austr., 36, e028
Fakhouri O., Ma C.-P., Boylan-Kolchin M., 2010, MNRAS, 406, 2267
Fernández-Trincado J. G., Forero-Romero J. E., Foex G., Verdugo T., Motta V., 2014, ApJ, 787, L34
Forero-Romero J. E., Gottlöber S., Yepes G., 2010, ApJ, 725, 598
Gastaldello F. et al., 2014, MNRAS, 442, L76
Golovich N., Dawson W. A., Wittman D., Ogrean G., van Weeren R., Bonafede A., 2016, ApJ, 831, 110
Golovich N., van Weeren R. J., Dawson W. A., Jee M. J., Wittman D., 2017, ApJ, 838, 110
Golovich N. et al., 2019a, ApJS, 240, 39
Golovich N. et al., 2019b, ApJ, 882, 69
Gunn J. E., Gott J., Richard I., 1972, ApJ, 176, 1
Ha J.-H., Ryu D., Kang H., 2018, ApJ, 857, 26
Harvey D., Massey R., Kitching T., Taylor A., Tittley E., 2015, Science, 347, 1462
Hayashi E., White S. D. M., 2006, MNRAS, 370, L38
Jee M. J., Mahdavi A., Hoekstra H., Babul A., Dalcanton J. J., Carroll P., Capak P., 2012, ApJ, 747, 96
Jee M. J., Hughes J. P., Menanteau F., Sifón C., Mandelbaum R., Barrientos L. F., Infante L., Ng K. Y., 2014, ApJ, 785, 20
Jiang L., Helly J. C., Cole S., Frenk C. S., 2014, MNRAS, 440, 2115
Kahlhoefer F., Schmidt-Hoberg K., Frandsen M. T., Sarkar S., 2014, MNRAS, 437, 2865
Kazantzidis S., Magorrian J., Moore B., 2004, ApJ, 601, 37
Khochfar S., Burkert A., 2006, A&A, 445, 403
Kim S. Y., Peter A. H. G., Wittman D., 2017, MNRAS, 469, 1414
Knebe A. et al., 2011, MNRAS, 415, 2293
Knebe A. et al., 2013, MNRAS, 435, 1618
Kraljic D., Sarkar S., 2015, J. Cosmol. Astropart. Phys., 2015, 050
Kravtsov A. V., Borgani S., 2012, ARA&A, 50, 353
Lage C., Farrar G., 2014, ApJ, 787, 144
Lage C., Farrar G. R., 2015, J. Cosmol. Astropart. Phys., 2015, 038
Lee J., Komatsu E., 2010, ApJ, 718, 60
Lewis A., Challinor A., Lasenby A., 2000, ApJ, 538, 473
Lourenço A. C. C. et al., 2020, MNRAS, 498, 835
Ludlow A. D. et al., 2013, MNRAS, 432, 1103
Machado R. E. G., Monteiro-Oliveira R., Lima Neto G. B., Cypriano E. S., 2015, MNRAS, 451, 3309
Mahdavi A., Hoekstra H., Babul A., Balam D. D., Capak P. L., 2007, ApJ, 668, 806
Markevitch M., Vikhlinin A., 2007, Phys. Rep., 443, 1
Markevitch M., Gonzalez A. H., David L., Vikhlinin A., Murray S., Forman W., Jones C., Tucker W., 2002, ApJ, 567, L27
Markevitch M., Gonzalez A. H., Clowe D., Vikhlinin A., Forman W., Jones C., Murray S., Tucker W., 2004, ApJ, 606, 819
Mastropietro C., Burkert A., 2008, MNRAS, 389, 967
Milosavljević M., Koda J., Nagai D., Nakar E., Shapiro P. R., 2007, ApJ, 661, L131
Monteiro-Oliveira R., Cypriano E. S., Vitorelli A. Z., Ribeiro A. L. B., Sodré L. J., Dupke R., Mendes de Oliveira C., 2018, MNRAS, 481, 1097
Moura M. T., Machado R. E. G., Monteiro-Oliveira R., 2021, MNRAS, 500, 1858
Navarro J. F., Frenk C. S., White S. D. M., 1996, ApJ, 462, 563
Ng K. Y., Dawson W. A., Wittman D., Jee M. J., Hughes J. P., Menanteau F., Sifón C., 2015, MNRAS, 453, 1531
Obreschkow D., 2022, gadgetry: Reading, Processing and Visualising Astrophysical N-Body Simulation Data. Available at GitHub, https://github.com/obreschkow/gadgetry.git
Obreschkow D., Elahi P. J., Lagos C. d. P., Poulton R. J. J., Ludlow A. D., 2020, MNRAS, 493, 4551
Peebles P. J. E., 1965, ApJ, 142, 1317
Pillepich A. et al., 2018, MNRAS, 473, 4077
Planck Collaboration XIII, 2016, A&A, 594, A13
Poole G. B., Fardal M. A., Babul A., McCarthy I. G., Quinn T., Wadsley J., 2006, MNRAS, 373, 881
Poulton R., 2019, Astrophysics Source Code Library, record ascl:1911.019
Predehl P., Andritschke R., Arefiev V., et al., 2021, A&A, 647, A1
R Core Team, 2022, R: A Language and Environment for Statistical Computing. R Foundation for Statistical Computing, Vienna, Austria, available at https://www.R-project.org/
Randall S. W., Markevitch M., Clowe D., Gonzalez A. H., Bradač M., 2008, ApJ, 679, 1173
Ricker P. M., Sarazin C. L., 2001, ApJ, 561, 621
Robotham A., 2022, magicaxis: Pretty Scientific Plotting with Minor-Tick and Log Minor-Tick Support. Available at https://CRAN.R-project.org/package = magicaxis
Roettiger K., Burns J. O., Loken C., 1996, ApJ, 473, 651
Sarazin C. L., 2002, in Feretti L., Gioia I. M., Giovannini G., eds, Astrophysics and Space Science Library, Vol. 272, Merging Processes in Galaxy Clusters. Springer, Berlin, p. 1
Schaye J. et al., 2015, MNRAS, 446, 521
Springel V., 2005, MNRAS, 364, 1105
Springel V., Farrar G. R., 2007, MNRAS, 380, 911
Takizawa M., 2005, ApJ, 629, 791
Takizawa M., 2006, PASJ, 58, 925
Thompson R., Nagamine K., 2012, MNRAS, 419, 3560
Thompson R., Davé R., Nagamine K., 2015, MNRAS, 452, 3030
Tucker W. H., Tananbaum H., Remillard R. A., 1995, ApJ, 444, 532
Tucker W. et al., 1998, ApJ, 496, L5
Tulin S., Yu H.-B., 2018, Phys. Rep., 730, 1
Van Weeren R. J. et al., 2017, Nat. Astron., 1, 0005
Watson W. A., Iliev I. T., Diego J. M., Gottlöber S., Knebe A., Martínez-González E., Yepes G., 2014, MNRAS, 437, 3776
White S. D. M., Frenk C. S., 1991, ApJ, 379, 52
White S. D. M., Rees M. J., 1978, MNRAS, 183, 341
ZuHone J. A., 2011, ApJ, 728, 54

# APPENDIX A: SMALL NUMBER EFFECTS ON DISSOCIATION

Haloes with a low number of particles can be subject to fluctuations in $S$ due to the Poisson noise associated with the particle sampling. Fig. A1 showcases two examples of poorly resolved haloes found in L210N1024NR, highlighting the extreme range of dissociation found in these poorly resolved haloes ($< 7.5 \times 10^{11}$ M$_\odot$, or $\lesssim$ 1000 particles per species) compared to the larger distribution of $S$ discussed in Section 3. By resampling contrived density fields with varying the number of particles, we found that of order $10^3$ particles per halo are needed for $S$ to be reasonably converged (to about a per cent).

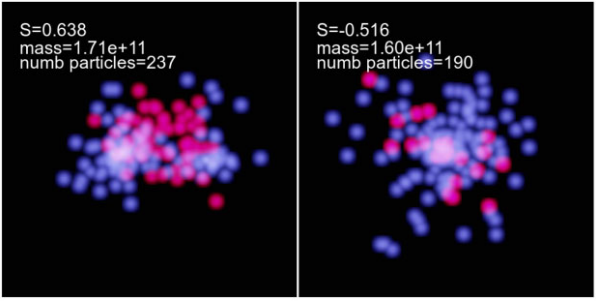

**Figure A1.** An illustration of two poorly resolved haloes from SURFS L210N1024NR, where from the low particle count, 237 (left-hand panel) and 190 (right-hand panel) the spatial distributions are of the ICM (red) and DM (blue) are under sampled leading to a more extreme *S*.

# APPENDIX B: PARAMETRIZATION OF TWO-BODY ORBITS

This section is a reminder of the general parametrization of two-body orbits.

For convenience, the first body of the two-body system is taken to be the more massive one, such that the mass ratio $\beta = M_2/M_1$ is bound to $\beta \in [0, 1]$. Any Keplerian orbit of the two masses can be characterized by six orbital elements. Of those, four can be ignored when studying halo–halo collisions, since we do not care about the orientation of the orbital plane (two parameters), the rotation of the Keplerian orbit inside this plane (one parameter), and the reference time (one parameter).

The two remaining degrees of freedom define the size and shape of the orbits. We can quantify them, for instance, by the pericentre distance $r_p$ and the eccentricity $e$. The latter can describe the full range from circular ($e = 0$), to elliptical ($0 < e < 1$), parabolic ($e = 1$), hyperbolic ($1 < e < \infty$), and rectilinear ($e = \infty$) orbits. A common choice is to use the orbital energy $E$ and angular momentum $L$ instead of $e$ and $r_p$.

In the context of two orbiting haloes, it is often convenient to normalize orbital parameters to the scale of a halo, for example the more massive one. Explicitly, we define the dimension-free periapsis distance $\hat{r}_p \equiv r_p/r_{hm,1}$, where $r_{hm,1}$ is the half-mass radius of the more massive halo. Likewise, we can express $E$ and $L$ as dimension-free parameters. Following Binney & Tremaine (1987), we choose them to be $\hat{E} \equiv 2Er_{hm,1}/(GM_1^2)$ and $\hat{L} = L/(GM_1^3 r_{hm,1})^{1/2}$.

The resulting parameter transformation $(\hat{E}, \hat{L}) \mapsto (e, \hat{r}_p)$ is

$$e = \sqrt{1 + \frac{2EL^2}{\mu(GM_1M_2)^2}} = \sqrt{1 + \hat{E}\hat{L}^2(1+\beta)\beta^{-3}}, \quad (B1)$$

$$\hat{r}_p = \frac{L^2}{(1+e)\mu GM_1M_2 r_{hm,1}} = \frac{\hat{L}^2(1+\beta)}{(1+e)\beta^2}, \quad (B2)$$

where $\mu = M_1M_2/(M_1+M_2)$ is the reduced mass. The corresponding inverse relations are

$$\hat{E} = \frac{\beta(e-1)}{\hat{r}_p}, \quad (B3)$$

$$\hat{L} = \sqrt{\frac{\beta^2 \hat{r}_p(e+1)}{\beta+1}}. \quad (B4)$$

The mapping between these two parameter spaces is visualized by Fig. B1.

# APPENDIX C: ADDITIONAL NUMERICAL DETAILS OF THE IDEALIZED SIMULATIONS

## C1 Base halo density profile

The density profile of the base halo discussed in Section 4.1 after it was evolved in a vacuum is shown in Fig. C1 to a radius of $R = r_{200}$ with the contributions from the gaseous (red) and dark matter (blue) components shown in comparison to the initial fitted NFW density profile (Navarro et al. 1996) (dashed black curve). When placed in a vacuum the base progenitor diverged from the sampled NFW profile at small radii and continued to evolve before becoming relatively stable after an extended period of time. We cannot say that the density profile is stable as the halo itself is not at equilibrium due to the particles being assigned initial velocities sampled from a local Gaussian distribution matched to the circular velocities. This result is predicted by Kazantzidis et al. (2004), a similarly constructed halo relaxed to a state where the inner regions of the density profile diverged from the NFW profile. Due to this sampling the halo initially expands and collapses, leading to a compressive wave that oscillates throughout the halo, creating disturbances in the gaseous components as seen in Fig. C1 by the sudden change in the slope of the gas density profile at $R/R_{200} \sim 0.02$ and $R/R_{200} \sim 0.01$. The extent that our final density profile diverges from the NFW profile is much greater than reported by Kazantzidis et al. (2004) due to the extended period of time that we have allowed the progenitor to evolve and ensure a 'stable' state is reached. Despite the deviations from the NFW profile shown in the inner regions of the base progenitor, we find the remaining outer regions ($R/R_{200} > 0.1$) are in reasonable agreement with the NFW profile when truncated to a distance of $R = 2r_{200}$. A linear correlation of 0.96 is seen between the evolved density profile and the initial NFW profile when they are mapped against one another at the same radii.

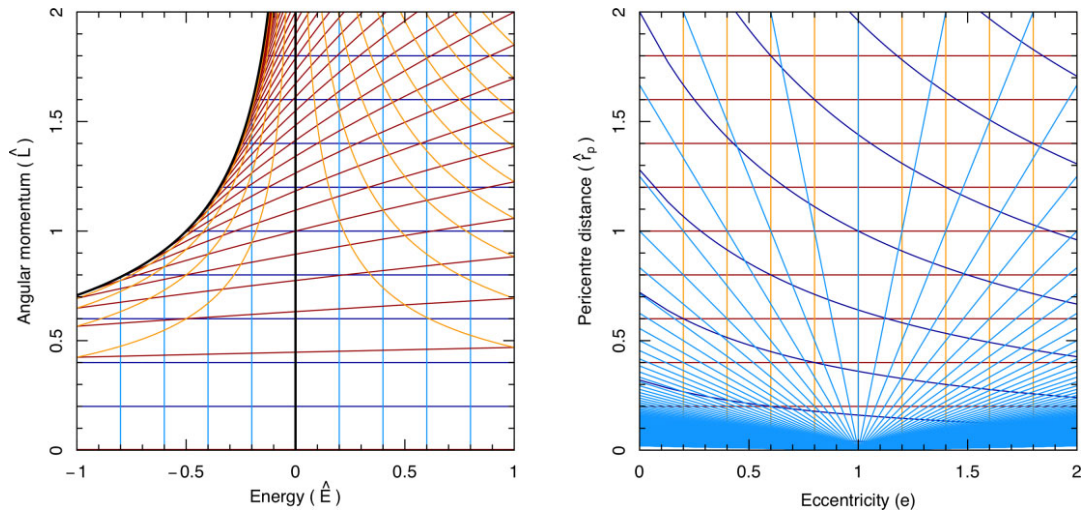

**Figure B1.** An illustration of the non-linear relationship between the two spaces $(e, \hat{r}_p)$ and $(\hat{E}, \hat{L})$ used to characterize binary orbits, when $\beta = 1$. The transformation of one space to another is given by equations (B1)–(B2). Left-hand panel: The normalized orbital energy ($\hat{E}$, light blue) and angular momentum ($\hat{L}$, dark blue) parameter space provided in Binney & Tremaine (chapter 7.4 1987). Right-hand panel: A parameter space defined by orbital eccentricity ($e$, orange) and a normalized pericentre distance ($\hat{r}_p$, red). In the left-hand panel the solid black curve depicts circular orbits ($\hat{L} = \hat{L}_{\rm circ}(\hat{E})$), it is not possible for an orbit to occur in the blank region above this line as circular orbits have the maximum possible angular momentum for a given $\hat{E}$ value (Binney & Tremaine 1987).

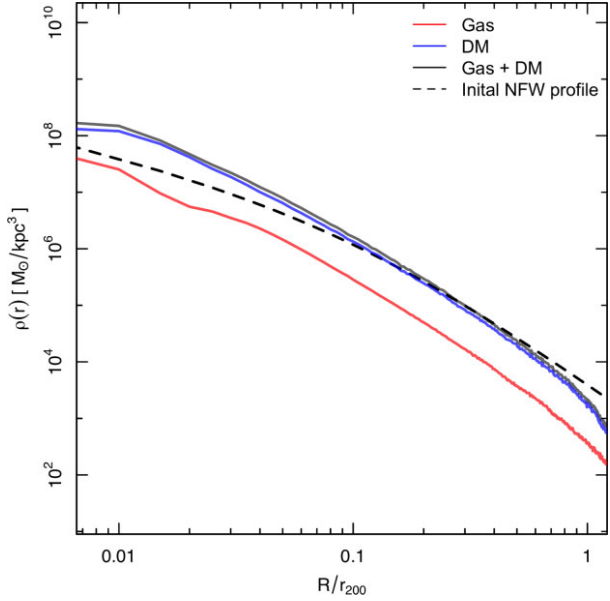

**Figure C1.** The density profile of the base halo, evolved in a vacuum for 12 Gyr (as discussed in Section 4). An NFW profile, which the base halo was initially sampled from is overlaid (dashed black curve) for comparison.

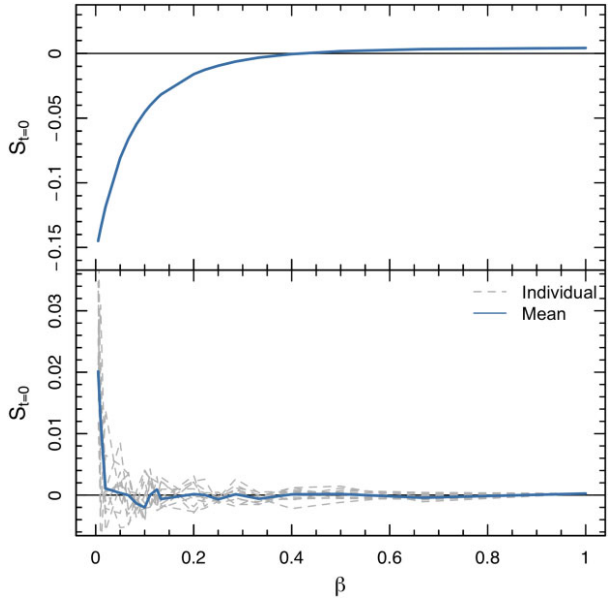

**Figure C2.** Top: The average dissociation at $t = 0$ ($S_{t=0}$), for two separated haloes as one is scaled in mass with respect to the other. Bottom: The dissociation of an isolated halo as it is scaled to various mass ratios.

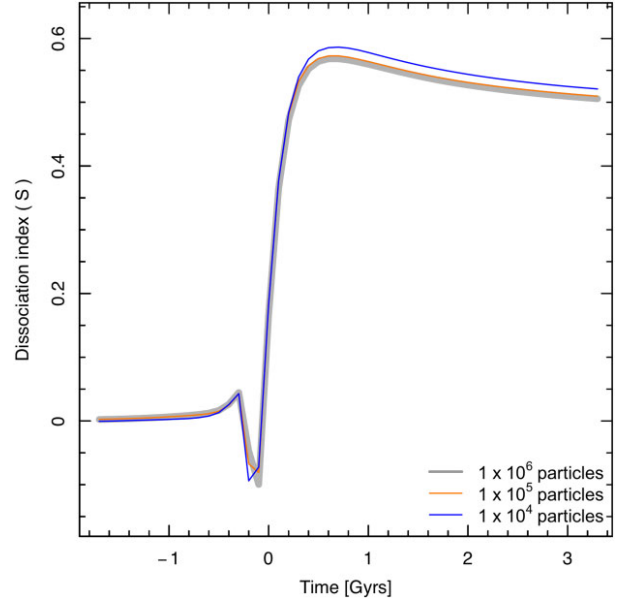

**Figure C3.** The profile of $S$ as a function of time for orbits characterized by $(e, \hat{r}_p, \beta) = (4, 0.75, 1)$ with progenitors each containing $1 \times 10^6$ (black), $1 \times 10^5$ (orange), and $1 \times 10^4$ (blue) particles.

To make one further addition to this discussion, both progenitors share the same density profile, therefore, we do not expect any significant impact on our results. It has been shown that it is the differences between the density profiles of progenitors that is impactful on dissociation (e.g. Moura et al. 2021).

### C2 Dissociation indices of initial conditions

The initial negative dissociation seen for orbits of $\beta = 1/10$ in Fig. 12 is a result of small asymmetries between the DM and gas within the base halo that are amplified by a reduced number of particles in the second (less massive) halo as it is scaled in accordance with $\beta$. To examine the stability of $S$ within the initial conditions of each idealized simulation at various mass ratios we measured $S$ of the progenitors at $t = 0$ ($S_{t=0}$) as a function of the mass ratio, $\beta$. Starting at $\beta = 1$ we scale one of the progenitors as outlined in Section 4.1.1, the result of which is presented in the top panel of Fig. C2. It can be seen when $\beta = 1/10$ that $S_{t=0} \approx -0.045$, which is the mean $S_{t=0}$ for such orbits presented in Fig. 12. Implying the offset is an artefact of scaling the base halo by randomly removing equal numbers of both ICM and DM particles. This test was repeated multiple times for just a singular halo in a vacuum, with the results shown in the bottom panel of Fig. C2.

In testing the impact of this initially negative dissociation, specific binary orbits were repeated with a purposefully sampled base halo where the initial dissociation is $S = 0$. No systematic differences in the behaviour of $S$ throughout the collision were observed for orbits characterized by $\hat{r}_p < 1$ or for orbits where a noticeable structural change occurred in either progenitor. A minor systematic correction can be made for all other orbits characterized by $\beta = 1/10$. However, this correction has a negligible impact on the results of this paper, given the affected orbits are not frequently occurring nor strongly dissociated.

### C3 Dissociation index convergence

To examine the relationship between *S* and the mass resolution of our ideal simulations we complete three control runs using two identical progenitors sampled by $1 \times 10^6$, $1 \times 10^5$, and $1 \times 10^4$ particles. The orbit is characterized by $(e, \hat{r}_p, \beta) = (4, 0.75, 1)$, which was randomly sampled from orbits with the potential to be strongly dissociative (see Fig. 13). The results of each control run are shown in Fig. C3.

If we consider the simulation with progenitors of $1 \times 10^6$ particles to have a sufficient level of accuracy in determining the dissociation index. Both runs with progenitors of $1 \times 10^5$ and $1 \times 10^4$ particles yield reasonably similar results, confirming that each idealized simulation will have a sufficient resolution. This result is unsurprising given that we find reasonable accuracy in the well-resolved haloes of SURFS L210N1024NR when they contain $\gtrsim 1000$ particles per species (see Section 3 and Appendix A).

## APPENDIX D: ADDITIONAL RESULTS

Fig. D1 contains the profile of *S* as a function of the snapshot number for each idealized binary collision as depicted in Fig. 11. The characteristic profile discussed in Section 4.3 is observed in the initial collision of each run. Within the profiles of elliptical and parabolic orbits that eventuate in a merger, following the

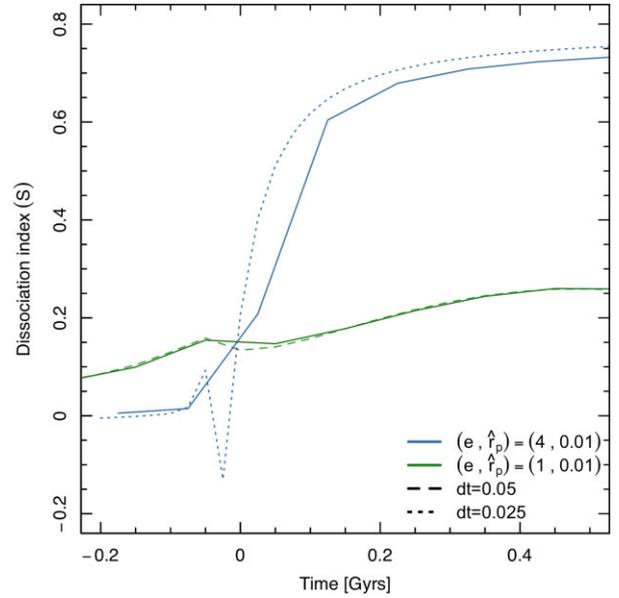

**Figure D2.** The profile of *S* for two orbits characterized by $(e = 4, \hat{r}_p = 0.01)$ (blue) and $(e = 1, \hat{r}_p = 0.01)$ (green). An initial decrease in *S* is not observed with a time step size of $t = 0.1$ Gyr (solid curve) for both orbits, however at a time step size of $t = 0.025$ (dotted) and $t = 0.05$ Gyr (dashed) this feature was recovered.

initial peak, there are obvious fluctuations of $S(t)$. In these orbits, each consecutive pericentric and apocentric passage is indicated by the continued fluctuations in $S(t)$. Following a collision where the progenitors retain a positive orbital energy and in the wake

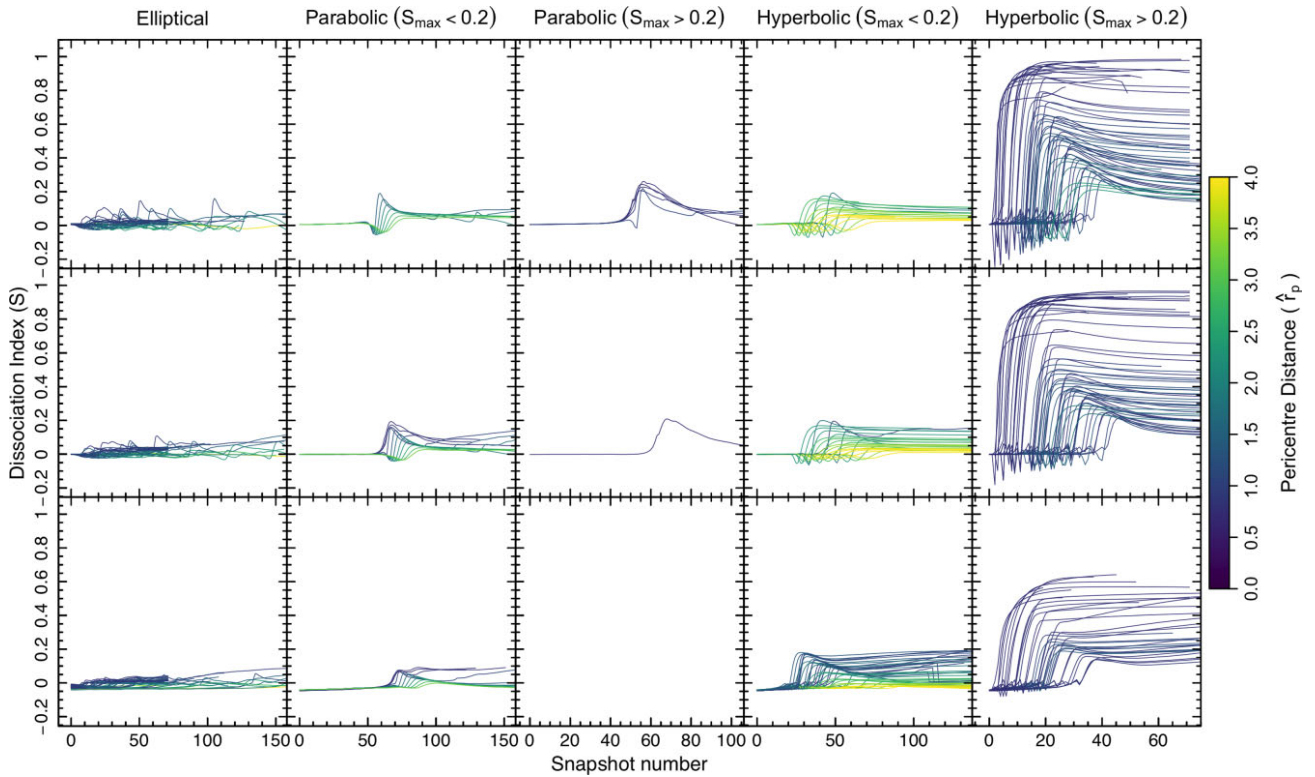

**Figure D1.** The evolution of *S* within each idealized binary collision simulated. Each of the three mass ratios sampled is divided into three rows (from top to bottom); $\beta = 1, 1/3, 1/10$. Each column of this figure is separated by the sampled eccentricity (elliptical ($e < 1$), parabolic ($e = 1$), and hyperbolic ($e > 1$)) and then again by the maximum dissociation observed ($S_{max}$). The pericentre distance ($\hat{r}_p$) is signified by colour.

of a merger $S$ asymptotically approaches a final value, i.e. we say $S$ 'relaxes'.

As noted in Section 4.3.2 and made apparent in Fig. D1 we observe a sub-population orbits with $\hat{r}_p \lesssim 0.1$ where an initial decrease in $S$ during the pericentric passage is not observed. We repeat the idealized simulations for a small sample of these orbits using smaller time intervals of $t = 0.05$ and $t = 0.025$ Gyr instead of the assumed $t = 0.1$ Gyr. In each instance we are able to recover this initial stagnation or decrease in $S$, implying it is an issue of sampling. These specific orbits are either highly energetic, with a high orbital velocity in which case the time in which this decrease occurs is too small to be consistently recovered or the decrease is minor, thus a smaller window in which it is detectable. Fig. D2 illustrates the former scenario with the hyperbolic orbit ($e = 4$, $\hat{r}_p = 0.01$) and the latter with a parabolic orbit ($e = 1$, $\hat{r}_p = 0.01$).

# APPENDIX E: ESTIMATING ORBITAL PARAMETERS

Within this paper we have applied a specified selection criteria on to the halo merger trees of L210N1536 to identify all collisions between host haloes, specifically to identify the orbits of haloes that lead to collisions as described in Section 5. Here we make note of the impacts of our selection criteria on the distributions of $e$ and $\hat{r}_p$.

## E1 Separation criterion

As described in Section 5.1 we identify progenitors at the onset of a collision between first-generation haloes and follow them through previous snapshots to where they are separated by a distance $R > \sum r_{hm}$. At which point the two progenitors are consistently identifiable as separate structures by the halo finder and the orbital parameters can be computed. A natural choice for this separation criterion is $R > \sum r_{200}$, however within our halo merger trees we experimentally determined that $\sum r_{hm,\,1}$ was more self-consistent as a function of redshift for a greater population of progenitor pairs than $\sum r_{200}$. Comparisons of the parameters estimated from each criterion show there is a negligible difference between the two, as shown in Fig. E1. Therefore we adopt $R > \sum r_{hm,\,1}$ over $R > \sum r_{200}$ for the greater sample size.

## E2 Uncertainty in estimated parameters

As explicitly stated within Khochfar & Burkert (2006) and echoed by Poulton (2019), determining the initial orbital parameters is a matter of timing. The progenitors cannot be distinguished from one another when they are too close together, as there is a large overlap, spatially, of mass between the two progenitors. Furthermore, environmental interactions can significantly alter the orbit and mass of either progenitor between snapshots. It is therefore possible to follow a progenitor back through too many snapshots, such that they can no longer be considered to be the same system or to have the same orbit. Hence the exact orbit and progenitors involved in the collision is not always recoverable. We, therefore, impose a criterion that the total mass of both progenitors cannot change by more than 10 per cent between snapshots (as done by Khochfar & Burkert 2006), to filter out orbits in overly dense environments in which neither progenitor can be identified or orbits affected by a major merger. However, this criterion allows for mass to be redistributed between the two progenitors by the halo finder when there is a significant overlap of the progenitors. Fluctuations of mass between the two progenitors imply the half mass radius of each will also fluctuate altering the time at which the required separation ($R > \sum r_{hm}$) is achieved and the precision with which the orbital parameters are estimated. To examine the uncertainty in the estimated orbital parameters we selected a population of binary collisions between first-generation haloes that occur at $z = 0$ and compare the estimated values when $R > \sum r_{hm}$ is achieved (Sn$_{est}$) with the estimated values at the snapshot prior (Sn$_{est} - 1$). The results are shown in Fig. E2, the linear regression fit of both $e$ and $\hat{r}_p$ show no systematic offsets between snapshots. From the regression fits we find a residual standard deviation of $\sigma_e = 0.129$ and $\sigma_{\hat{r}_p} = 0.099$.

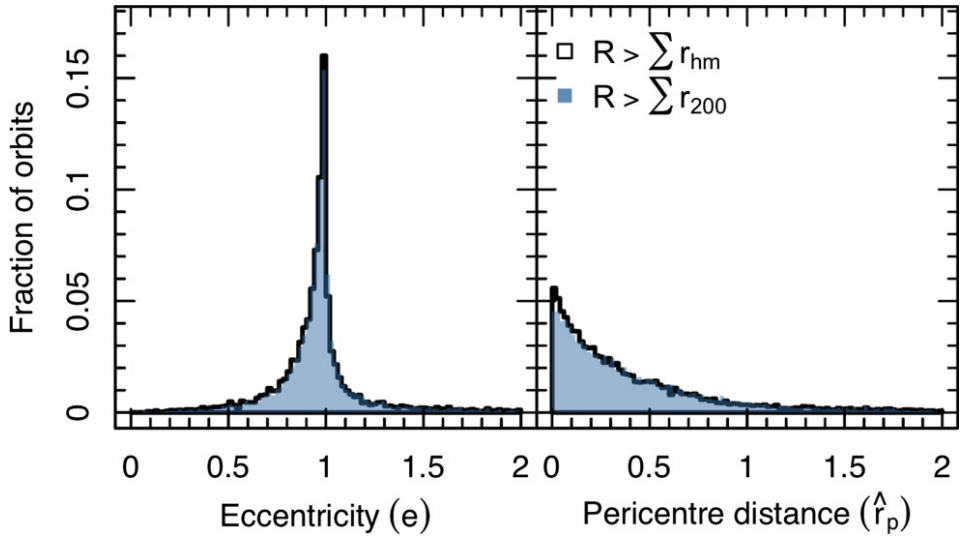

**Figure E1.** An illustration of the impact of selecting between $R > \sum r_{hm}$ (black outline) and $R > \sum r_{200}$ (blue shaded) as a criterion for the minimum separation on the distribution of orbital eccentricity (left-hand side) and normalized pericentre distance (right-hand panel).

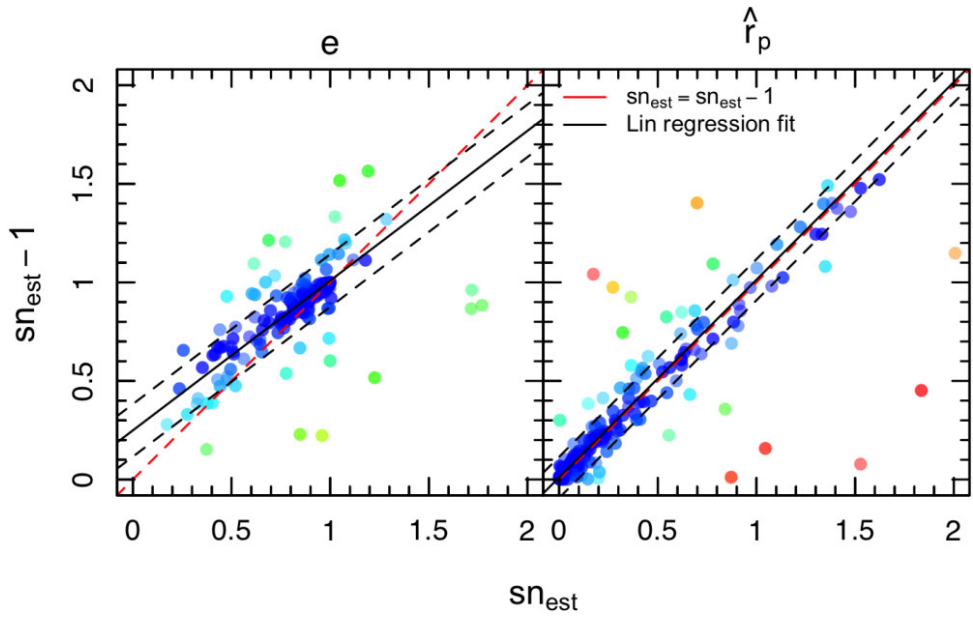

**Figure E2.** A comparison of the eccentricity $e$ (left-hand panel) and normalized pericentre distances $\hat{r}_p$ (right-hand panel) calculated when progenitors are first separated by a distance $R > \sum r_{hm}$ ($Sn_{est}$) with the prior snapshot ($Sn_{est} - 1$). The linear regression fit (solid line) with the residual standard deviation (dashed line) are shown in comparison to a 1:1 fit (red dashed line). Each point is coloured in accordance to the number of standard deviations from the the fitted linear regression line.

This paper has been typeset from a TEX/LATEX file prepared by the author.